\documentclass[12pt,preprint]{aastex}

\slugcomment{To appear in AJ (April 2003 issue)}

\shorttitle{Near-IR spectra of Cha I Stars}
\shortauthors{G\'omez and Mardones}

\begin{document}


\title{Near-infrared Spectra of Chamaeleon I Stars\altaffilmark{1}}


\author{M. G\'omez\altaffilmark{2}}
\affil{Observatorio Astron\'omico de C\'ordoba, 
    Laprida 854, 5000 C\'ordoba, Argentina}
\email{mercedes@oac.uncor.edu}

\and

\author{D. Mardones\altaffilmark{2}}
\affil{Departamento de Astronom\'\i a, Universidad de Chile, 
    Casilla 36-D, Santiago, Chile}
\email{mardones@das.uchile.cl}


\altaffiltext{1}{Based on observations collected at the European Southern
Observatory, Chile, (ESO proposal N.63.I-0269(A)).} 

\altaffiltext{2}{Visiting Astronomer, Cerro Tololo Inter-American Observatory.
CTIO is operated by AURA, Inc.\ under contract to the National Science
Foundation.}


\begin{abstract}
We present low resolution (R $\sim$ 500) near-infrared spectra of 46 candidate
young stellar objects in the Chamaeleon I star-forming region recently detected
in several deep photometric surveys of the cloud. Most of these stars have $K <$ 12.
In addition, we present spectra of 63 previously known southern hemisphere young
stars mainly belonging to the Chamaeleon I and Lupus dark clouds. We describe
near-infrared spectroscopic characteristics of these stars and use the
water vapor indexes to derive spectral types for the new objects.
Photometric data from the literature are used to estimate the bolometric
luminosities of all sources.  We apply D'Antona \& Mazzitelli (1998)
pre-main sequence evolutionary tracks and isocrones to derive masses and ages.   
We detect two objects with mass below the H burning limit among the 46 new candidates.
One of this object (PMK99 IR Cha INa1) is the likely driving source of a bipolar outflow in the
northern region of the cloud. 

Combining our targets with previously known members of the cloud 
we analyze the mass and age distributions for 145 stars in the Chamaeleon I
dark could. The mass histogram rises from about 2.5 ${\rm M\sun}$ up
to 0.4 ${\rm M\sun}$ and then falls off. The median mass is 0.30 ${\rm M\sun}$. 
The current population with masses $>$ 0.4 ${\rm M\sun}$ is essentially complete. 
The scarcity of very low mass members is interpreted as population 
bias towards the least massive and fainter objects. 
If we assume the {\it true} Chamaeleon I IMF is flat (in logarithmic mass bins) in the
interval 0.4 -- 0.04 ${\rm M\sun}$ as found by \citet{com00} 
in the central 300 arcmin$^2$ region, then we estimate that $\sim$ 100 stars
remain to be found in that mass range.  The distribution of ages indicates an
active star-formation episode within the last $\sim$ 5 $\times$ 10$^5$ yr and a
decreasing rate at older ages (few $\times$ 10$^7$ yr).
\end{abstract}




\section{Introduction}

In recent years there have been many infrared surveys of the Chamaeleon
I dark cloud \citep{cam98, per99, oas99, per00, goke01, per01}.  These
surveys provide many new candidate pre-main sequence stars 
based on photometric criteria. These 
observations aim to detect the faintest and thus the least massive pre-main
sequence stars and are, in most cases, sensitive enough to detect young
stars with masses close or even below the H burning limit. 
A complete census of the stellar population in the cloud  
would provide better estimates of the
mass distribution and the star-formation history of Chamaeleon I.  In
particular, these surveys open the door to study the IMF down to the substellar
regime.  The Chamaeleon I cloud is especially adequate for these purposes
since it is located nearby \citep[d $\sim$ 160 pc,][]{whi97}, while also
extending over moderate angular scales on the sky \citep[$\sim$ 3 deg$^{\rm
2}$,][]{bou98,miz01}.

\citet{law96} determined the IMF of the cloud down to $\sim$0.3 ${\rm
M\sun}$ based on $\sim$ 80 optically visible known members and found
a good agreement with the Miller-Scalo model.  \citet{com00} 
intensively observed a small region (300 arcmin$^2$) in the center of
Chamaeleon I dark cloud. This area contains 22 young stars, including 
13 very low mass members with K $\sim$ 11--13.5 identified in H$\alpha$. 
They found that the IMF for the central area of the Chamaeleon I
cloud was roughly flat (in logarithmic mass units)
from $\sim$ 1 ${\rm M\sun}$ down to $\sim$ 0.03 ${\rm M\sun}$. 

In order to study the extended low-mass IMF in the whole Chamaeleon I,
it is desirable to determine characteristics of the new
candidate young stars such as spectral types. The near-infrared
spectra combined with published pre-main sequence evolutionary
tracks can also be used to derive stellar masses and ages.  In this
contribution we present near-infrared spectra of 46 of the brightest
({\it typically} $K <$ 12) candidate young stellar objects selected
from the surveys cited above.  We chose the brightest objects in
order to have reasonable completeness and to achieve good
S/N ratios in 4-m class telescopes.

In addition, we obtained near-infrared spectra of 63 previously known
members of this cloud as well as other nearby southern star-forming regions
such as Lupus.  We used these known young stellar objects as a spectral
type calibration group. These known objects and 
the \citet{grla96} atlas allow us to identify 
spectroscopic features usually present in young stellar objects that support
the pre-main sequence status of the new objects.  We combine our data
with published photometry and spectral types of other sources in the cloud
to compile a list of 145 pre-main sequence stars in the Chamaeleon I dark cloud
and study the mass and age distributions.

In \S 2 we describe the observations and data reduction.  In \S 3 we present
our analysis and results. We derive spectral types from the water vapor
indexes which combined with published photometry allows us to place the
stars in the HR diagram. We conclude with a brief summary in \S 4.

\section{Observations and Data Reduction}

We carried out these observations on April 10 -- 12, 1999 
with the ESO NTT near-infrared spectrograph/imaging camera SOFI (Son OF ISAAC)
and on May 6 -- 9, 1999 with OSIRIS (Ohio State Infrared Imager and
Spectrometer) on the CTIO Blanco 4-m telescope. A few additional spectra
were obtained on Feb 27 --28, 2002 with this telescope and infrared
camera. Tables 1 and 2 list the targets observed. 

SOFI has two low resolution grisms (red and blue)
that roughly cover the $JHK$ bands on a Hawaii HgCdTe 1024$\times$1024
detector with a plate scale of 0.292$''$/pix.
The blue grism covers the spectral region between
0.95--1.63 $\mu$m and the red grism the region between
1.53--2.52 $\mu$m. The corresponding spectral resolutions
(R $=$ $\lambda$/$\Delta$$\lambda$) are 930 and 980 for a 0.6$''$ slit.
We used a 1$''$ wide and 290$''$ long slit. 
OSIRIS was used in the X-Disp (multi-order/cross-dispersed)
mode with the f/2.8 camera and a slit 30$''$ long and 1.2$''$ wide. 
This configuration gives R ($\lambda$/$\Delta$$\lambda$)
$\sim$ 1200 (2 pixels), while covering the $J$, $H$,
and $K$ bands simultaneously on a 1024 $\times$ 1024 HgCdTe
detector at a plate scale of 0.403$''$/pix\footnote{A full
description of this instrument and its various configurations
can be found at http://www.ctio.noao.edu/instruments/ir-instruments/osiris/index.html.}.
In particular, the X-Disp mode uses a cross-dispersing grism that gives
the $JHK$ spectral region in three adjacent orders, from 5th to 3rd. 

The integration time and the number of exposures per target
were chosen based on the brightness of the source
and background contribution each night. 
In most cases we obtained two sets of 9 blue and red spectra of the same
integration time with SOFI. We repeated this sequence for objects fainter
than $K =$ 13. Common integration times were
90--180 sec for both grisms.  The telescope was
nodded 30--60$''$ along the 290$''$ long slit between consecutive positions
following the usual ABBA pattern. This procedure corresponds to the
``Nod Throw Along Slit'' scheme as described in the SOFI Users Manual
\citep{lid00}. With OSIRIS we {\it typically} obtained seven images of identical
integration time, with the telescope dithered by 7$''$ along the 30$''$ long slit.
The individual integration times varied between 30 and 180 sec. For faint
objects (i.e., $K >$ 12) we obtained two 180 sec sequences of seven spectra.

In addition to the program sources we observed several
atmospheric standards. We selected two groups of standards
with similar airmass to our candidate sources and observed
them (every $\sim$ 2 hours) each night.
These telluric stars comprise both late (G3-5) and early (O8)
spectral types and were chosen from the lists available at both observing
sites. 

We obtained multiple flat field images, with a dome screen, using
incandescent lamps on and off. A Xenon lamp, also taken on and off, each
night provided the wavelength calibration for our SOFI data. To 
determine the wavelength scale of the OSIRIS spectra we used a HeNeAr lamp.

To reduce the data we used IRAF\footnote{IRAF is distributed by
the National Optical Astronomy Observatory, which is operated by
the Association of Universities for Research in Astronomy, Inc.
under contract to the National Science Foundation.}.     
We subtracted one image from another (using pairs of nodded observations) to
eliminate the background and sky contribution in first approximation.
This subtraction automatically took care of dark current and bias level.
We flat-fielded our data dividing by a normalized dome flat.   

We used the {\it twodspec} task APALL to trace and extract the
spectra along a 6 pixel wide aperture in the SOFI data and
10 pixel in the OSIRIS co-added images. In particular for OSIRIS data
we treated $JHK$ X-Disp adjacent orders using the {\it echelle}
package. A further sky subtraction was
done by fitting a polynomial to the regions on either
side of the aperture.  A non linear low order fit to the
lines in the Xenon and HeNeAr lamps were used to wavelength
calibrate the spectra.                  

We removed telluric features from our data dividing each program
star by the atmospheric standard spectrum. For the OSIRIS data
we used a G3-5 standard and both G3-5 and O8 stars for the SOFI spectra. The
telluric star differed from the science target by $<$ 0.2 in airmass.
We used the {\it onedspec} task SPLOT to remove H I (Paschen and Brackett)
lines from the spectra of the G3-5 and O8 stars, interpolating across each line.
The division by the standard canceled out
telluric features in the science spectrum reasonably well but
introduced the inverse of the atmospheric standard in
the corrected spectrum.  We recovered the true spectral shape
multiplying the resultant spectrum by a Planck function at
the temperature corresponding to the atmospheric star.               

We finally combined the $JHK$ X-Disp orders for our OSIRIS spectra and
the blue and the red grism regions for each object observed with SOFI
and eliminated regions of deep atmospheric absorptions. We also
trimmed out wavelengths 0.95--0.97 $\mu$m from our SOFI spectra due
to edge effects. The useful spectral range provided by OSIRIS is 12000-23500 \AA~ while SOFI 
covers from 9700 to 25300 \AA. For the SOFI spectra, as mentioned before,
we obtained two sets of spectra corresponding to the G3-5 and O8 telluric
stars that were essentially identical. In Section 3.2 we show   
the set of science spectra obtained applying the O8 standard as telluric
corrector. 

\section{Data Analysis and Results}

\subsection{The Sample}

We observed two groups of stars: a) 46 candidate young stellar objects
detected by several recent surveys of the cloud; and b) 63 previously
known young stars (36 belonging to the Chamaeleon I dark cloud and 27
located in other star-forming regions, such as Lupus). Three objects in 
the second group (VW Cha, SZ 119, and SZ 124) were observed with both
SOFI and OSIRIS. Eight of the stars in the second group already had
near-infrared spectra. \citet{grla96} observed Haro1-1 in the $\rho$ Ophiuchi cloud and
\citet{gope02} CCE98 49 (ISO-ChaI 256) in Chamaeleon I.  \cite{com99} published combined 
$HK$ band spectra for Cha H$\alpha$ 1--6.  \cite{gope02} obtained
individual $JHK$ band spectra for Cha H$\alpha$ 1 and Cha H$\alpha$ 2. 
In this paper we report individual $JHK$ data for each of the Cha H$\alpha$ 1--6 stars. 

The 46 new objects were selected, in general, among the brightest
({\it typically} K $<$ 12) near-infrared objects detected by several
recent surveys in the cloud \citep{cam98, per99, oas99, per00, goke01,
per01}. These lists include about $\sim$ 150 additional potential members of the
cloud with K $>$ 13. In this contribution we analyze data for the brightest
objects for which it is feasible to obtain near-infrared spectra with
a 3-4 meter class telescope in about 20-40 minutes total integration time. 

The second set of 63 previously known objects were selected among the southern 
hemisphere pre-main sequence stars 
to have a group of reference from which to derive common near-infrared spectroscopic features
among young stellar objects. Objects with previously known 
spectral types within this group allowed us to calibrate the
water vapor index ${\rm I_{H_2O}}$ based classification (see section 3.4).  

\subsection{The Near-infrared $H-K$ vs $J-H$ Diagram}

Tables 3 and 4 compile photometric and spectroscopic data (when available) for the observed
stars.  Figure 1 shows the position of 
new (stars) and previously known (dots) objects
in the near-infrared color-color
diagram.  Superposed on this
plot are the loci of unreddened main sequence dwarfs
\citep[solid line,][]{bebr88} and CTTS --classical T Tauri stars--
\citep[long-dash line,][]{mey97}. The reddening band \citep[dotted
line,][]{goke01} and the length of the reddening vector \citep{rile85}
are also indicated.      

\begin{figure}
\centering
\includegraphics[width=17cm]{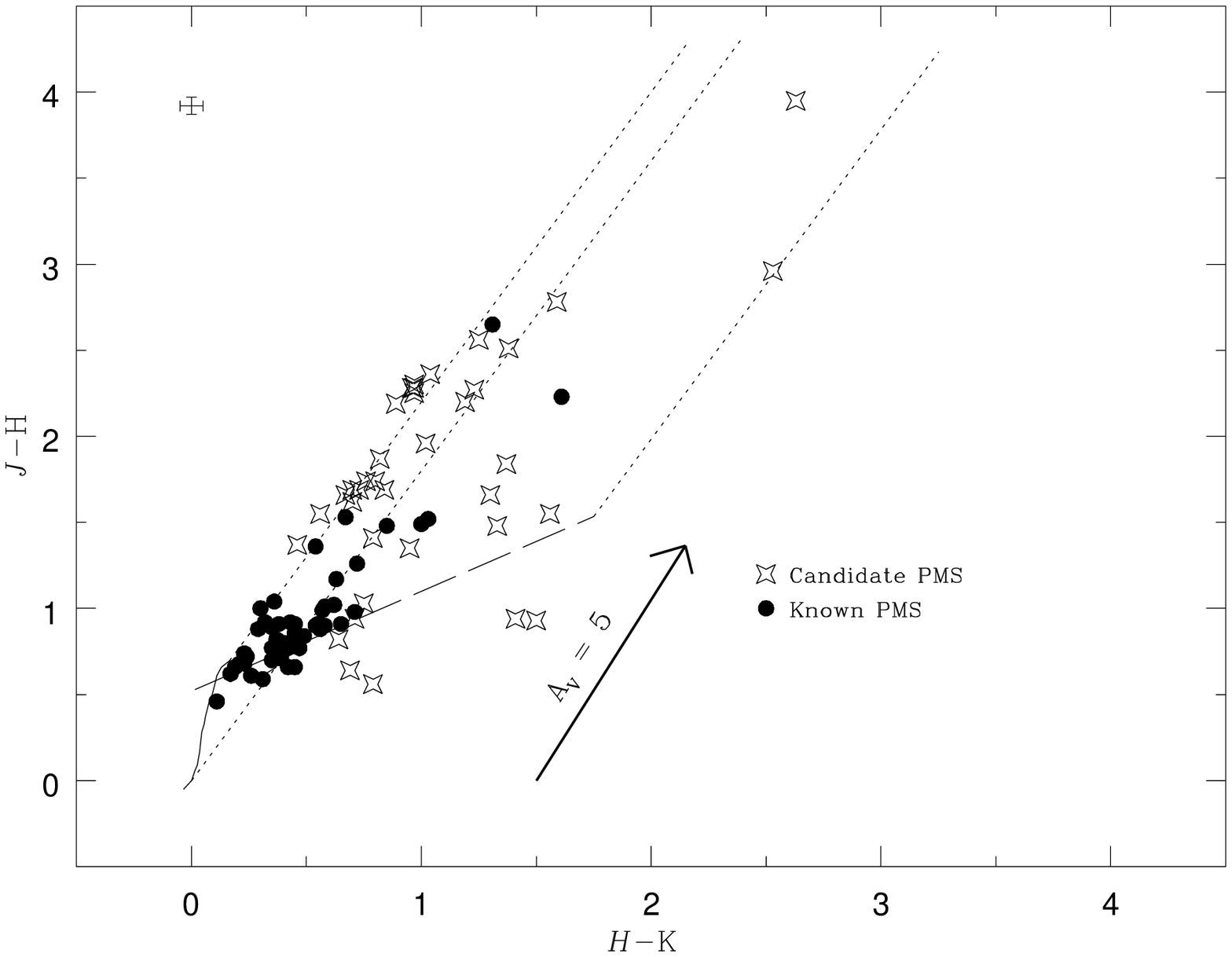}
\caption{Near-infrared color-color diagram for the candidate (stars)
and previously known (dots) sources listed in Tables 3 \& 4.  The solid and
long-dash lines indicate the loci of unreddened main sequence dwarfs
\citep{bebr88} and CTTS \citep{mey97}. The dotted lines define the
reddening band, corresponding to a reddening vector \hbox{E($J-H$)/E($H-K$)
$=$ 1.80} \citep{goke01}.  The arrow indicates an A$\rm _V$ $=$ 5 mag
\citep{rile85}. Typical photometric errors are displayed in the upper left
corner.}
\label{Fig1}
\end{figure}    

Although some of the candidate objects have significant near-infrared
excesses indicating the presence of circumstellar material, most of the
previously known and new sources are located within or close to the reddening band.  
The group of known objects includes many bona-fide CTTS in the cloud
\citep[e.g.,][]{gast92,com00} without measurable near-infrared excesses and thus
the near-infrared color-color diagram may miss a significant number of young
members of the cloud. Most of the candidate sources located
within the reddening band in Figure 1 were proposed by \citet{cam98} 
based on the $I-J$ vs $J-K$ diagram. Near-infrared spectra 
provide additional information to support the pre-main sequence nature
of these candidate sources. 

\subsection{The Spectra}

Figure 2 shows the $JHK$ band spectra of all candidate objects reported in this
paper and Figure 3 a selected sub-sample of 12 of the previously known young stellar
objects we observed \footnote{The spectra of the remanding 51 known stars as 
well as those shown in Figures 2 and 3 are available upon request to the
authors.}. Prominent atomic and molecular features in the
observed spectral range for late type stars (i.e., spectral types later than
G0) are labeled \citep{klha86, wal00, mey98, wahi97}. 
Both groups (new objects --Figure 2-- and previously known young 
stars --Figure 3--) have a wide variety of spectral shapes. Some
sources have shapes decreasing with $\lambda$, others display a turnover around
1.6--1.8 $\mu$m or the spectral shape is roughly constant.
Other objects have rising or steeply rising shapes towards 2 $\mu$m. 
Considered as groups, new young stellar objects and previously known
stars show similar near-infrared spectroscopic characteristics.  

\begin{figure}
\centering
\includegraphics[width=7.0in]{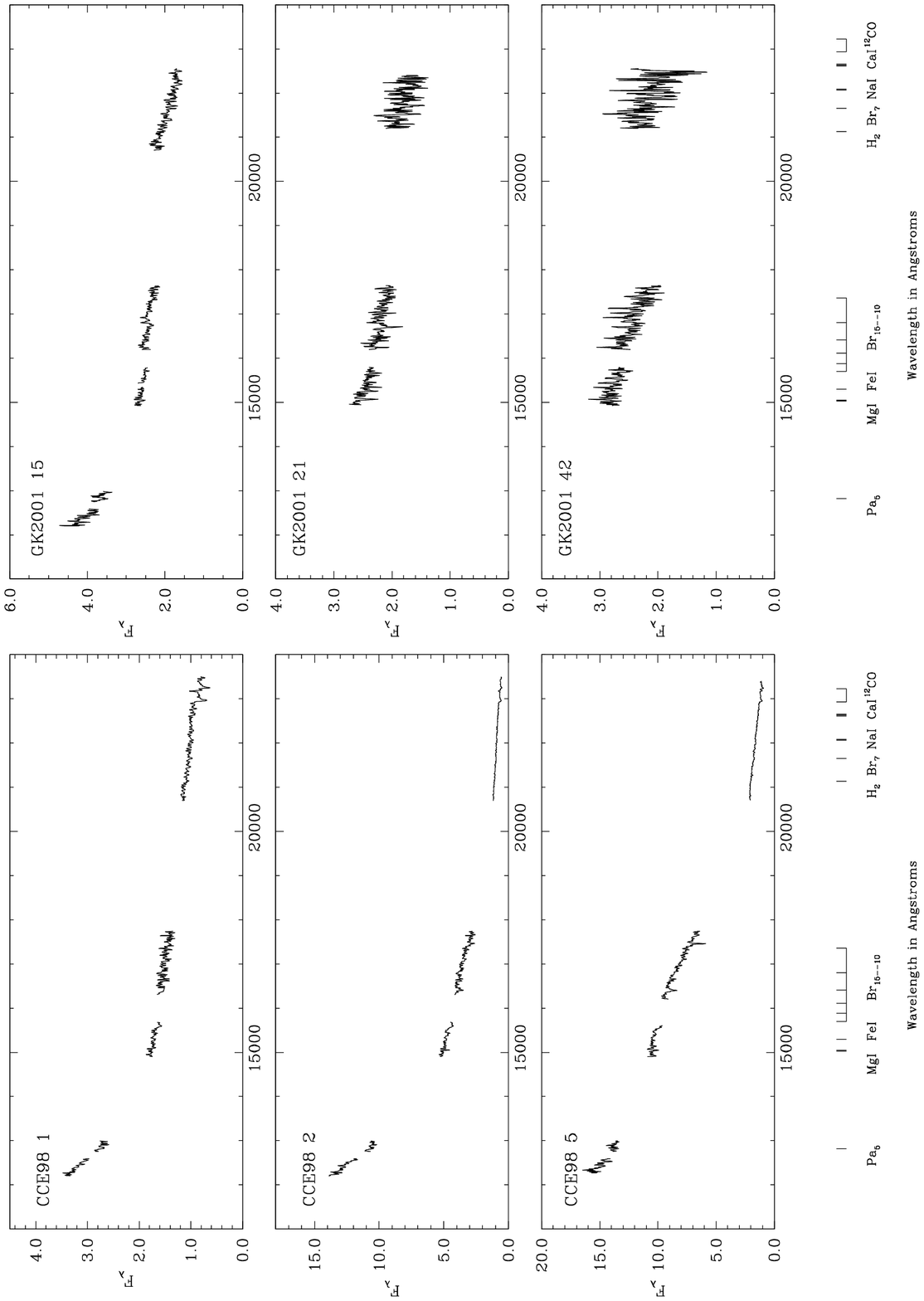}
\end{figure}       

\begin{figure}
\centering
\includegraphics[width=7.0in]{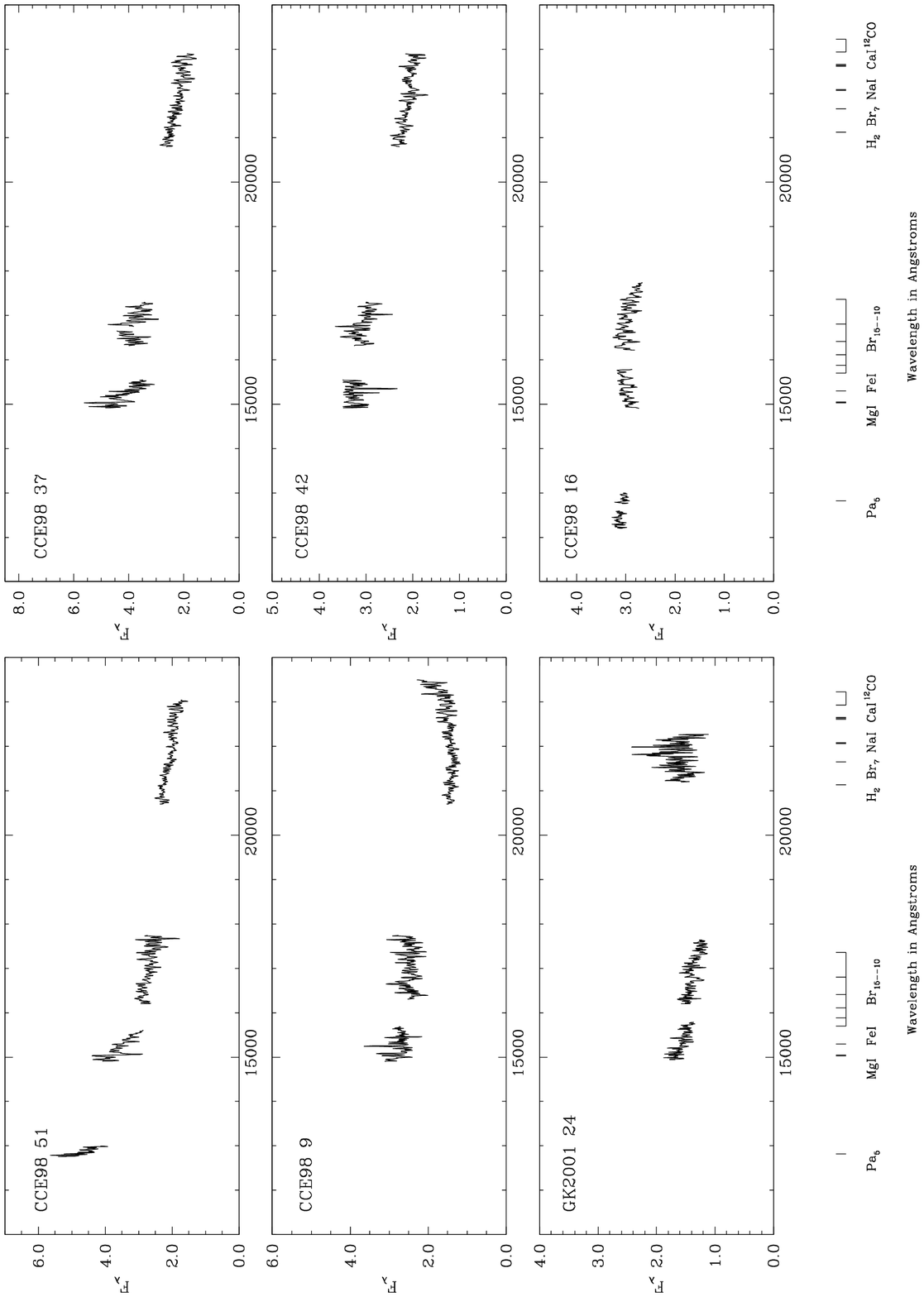}
\end{figure}

\begin{figure}
\centering
\includegraphics[width=7.0in]{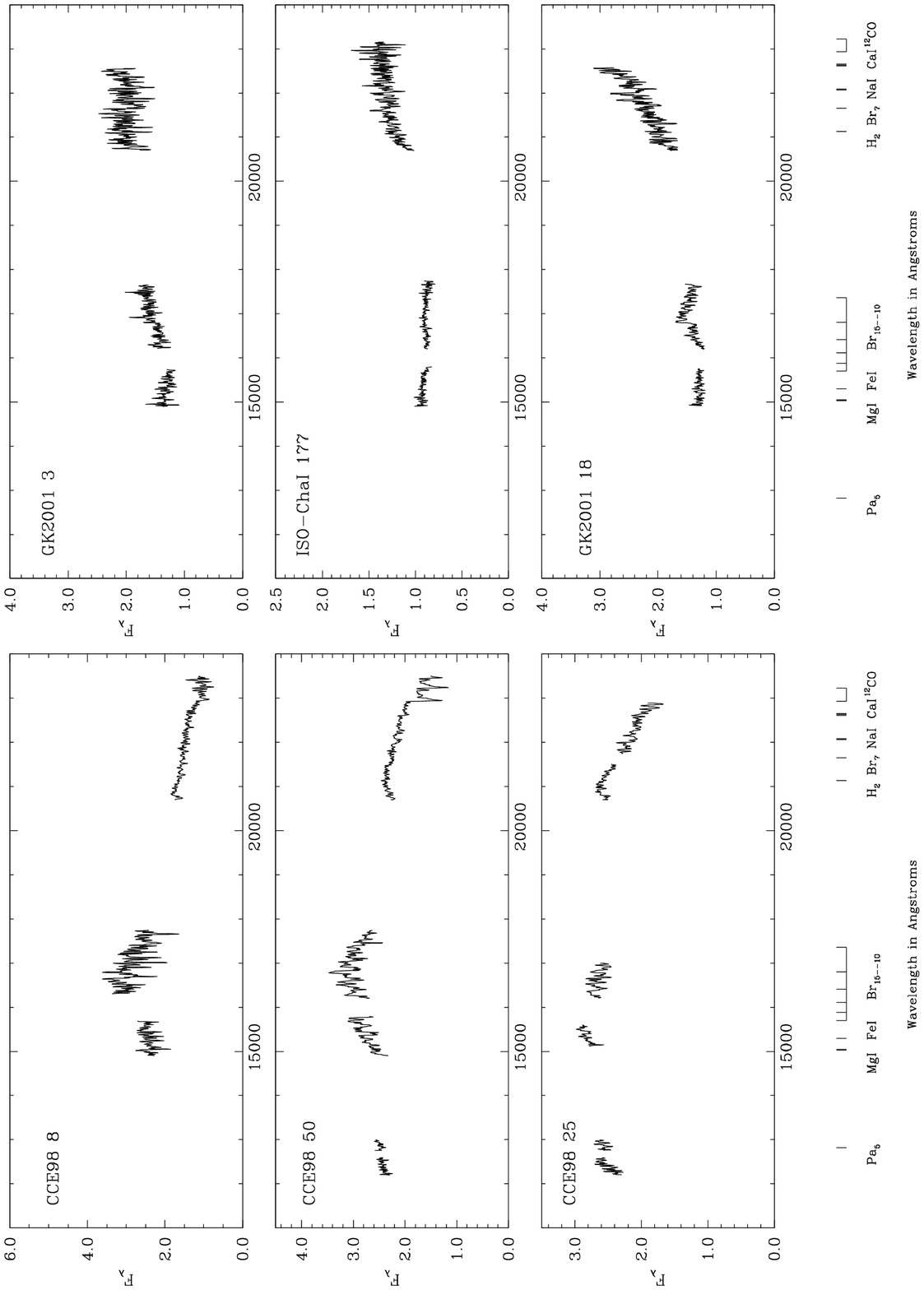}
\end{figure} 

\begin{figure}
\centering
\includegraphics[width=7.0in]{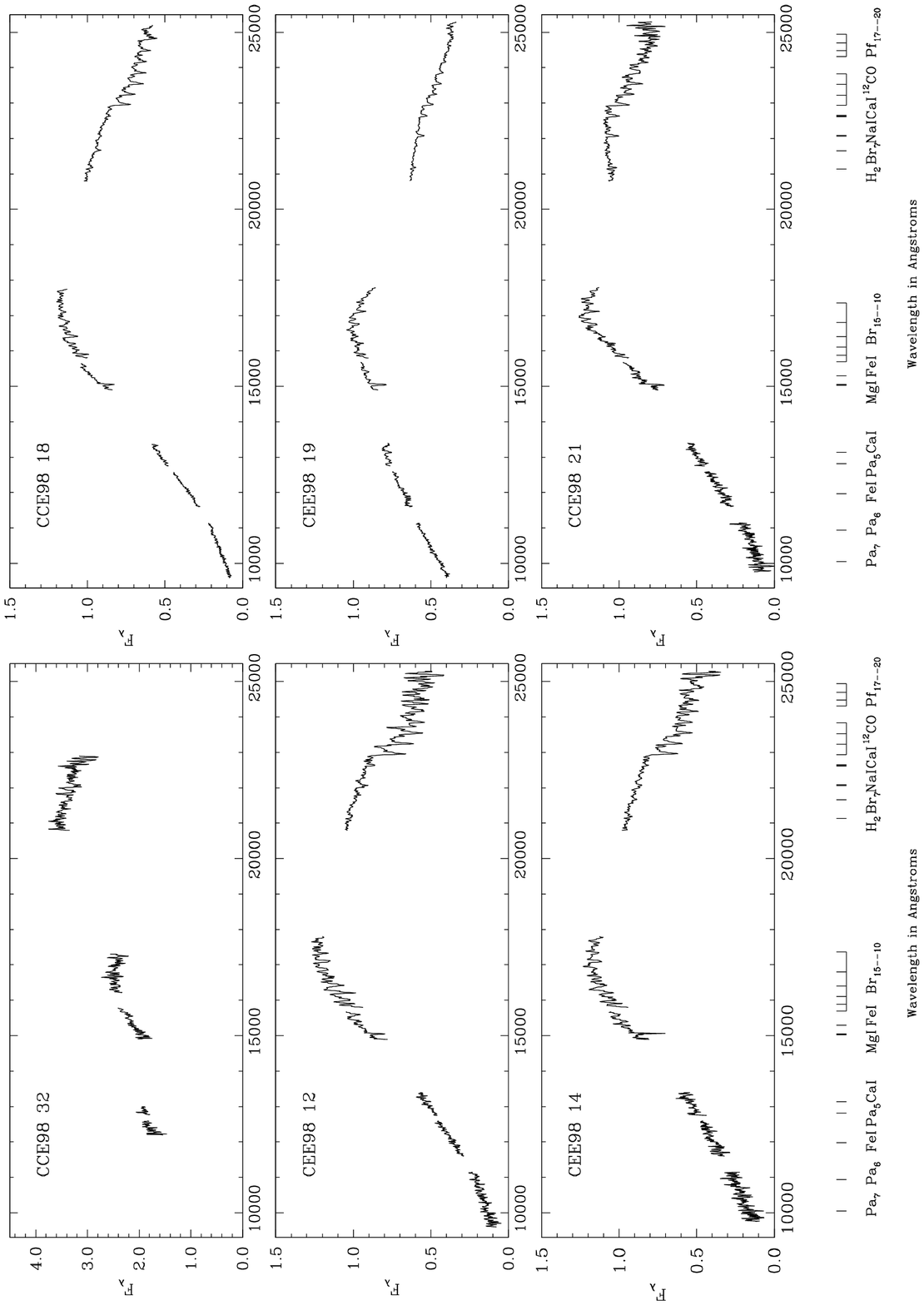}
\end{figure} 

\begin{figure}
\centering
\includegraphics[width=7.0in]{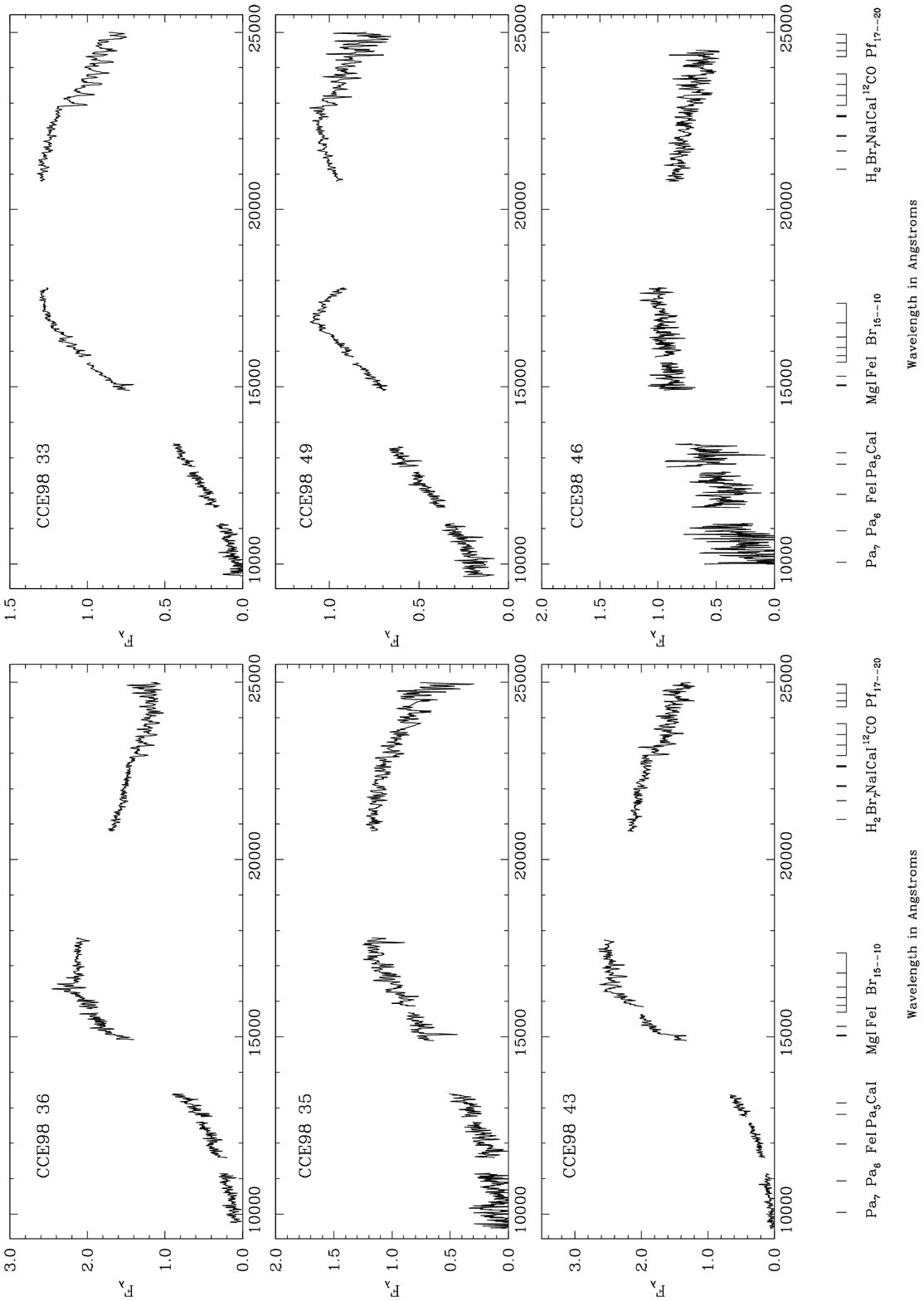}
\end{figure} 

\begin{figure}
\centering
\includegraphics[width=7.0in]{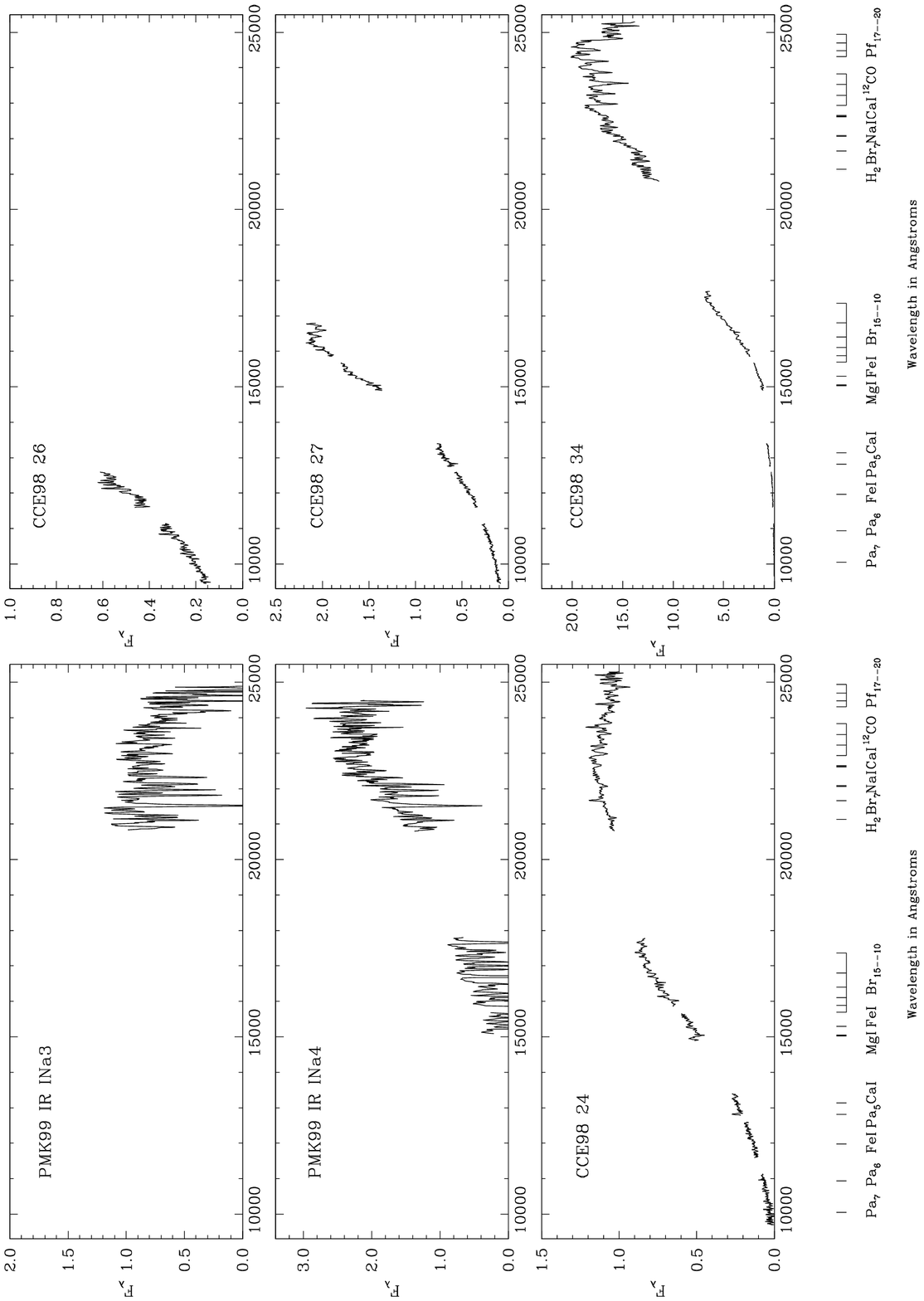}
\end{figure} 

\begin{figure}
\centering
\includegraphics[width=7.0in]{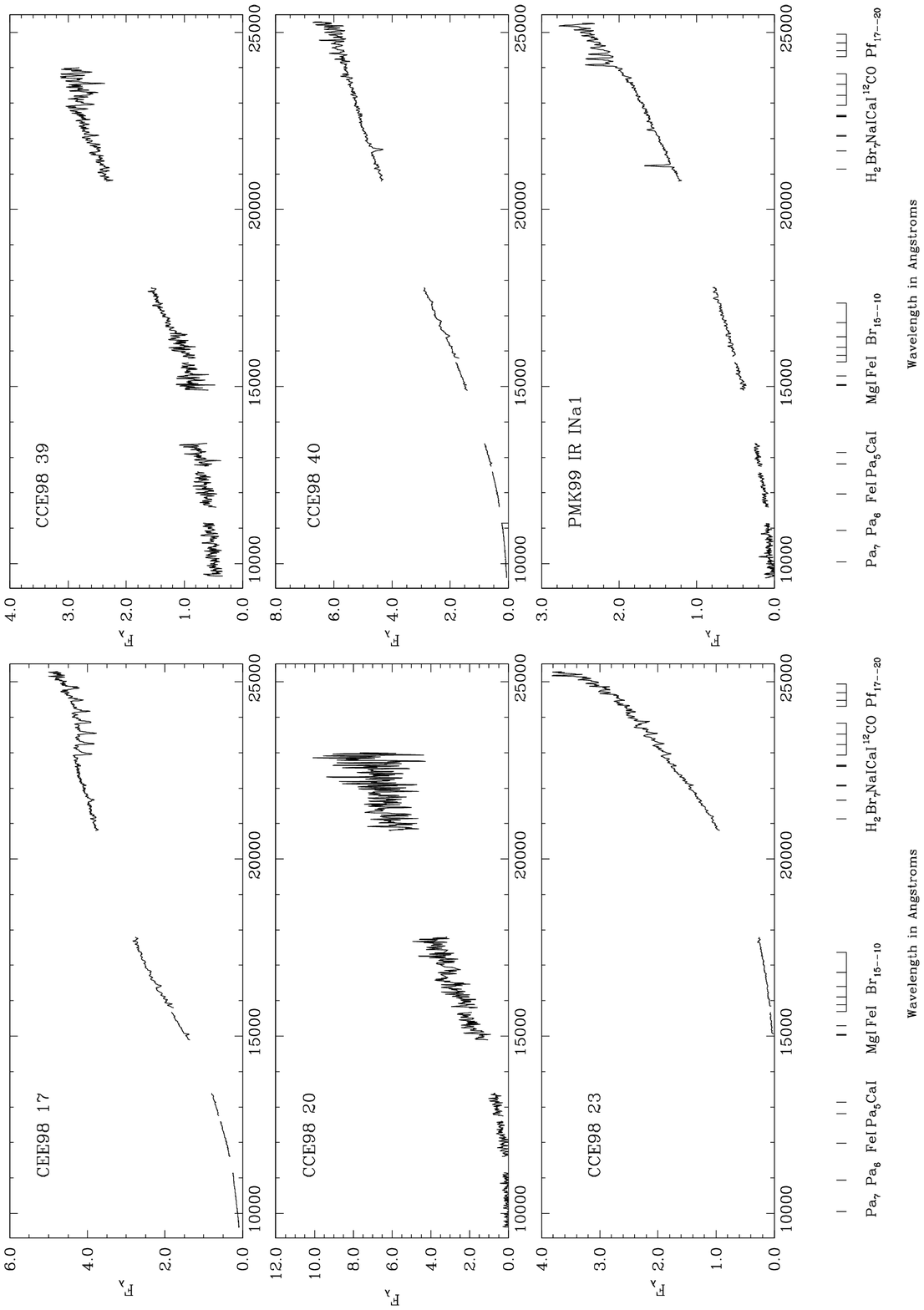}
\end{figure} 

\begin{figure}
\includegraphics[width=7.0in]{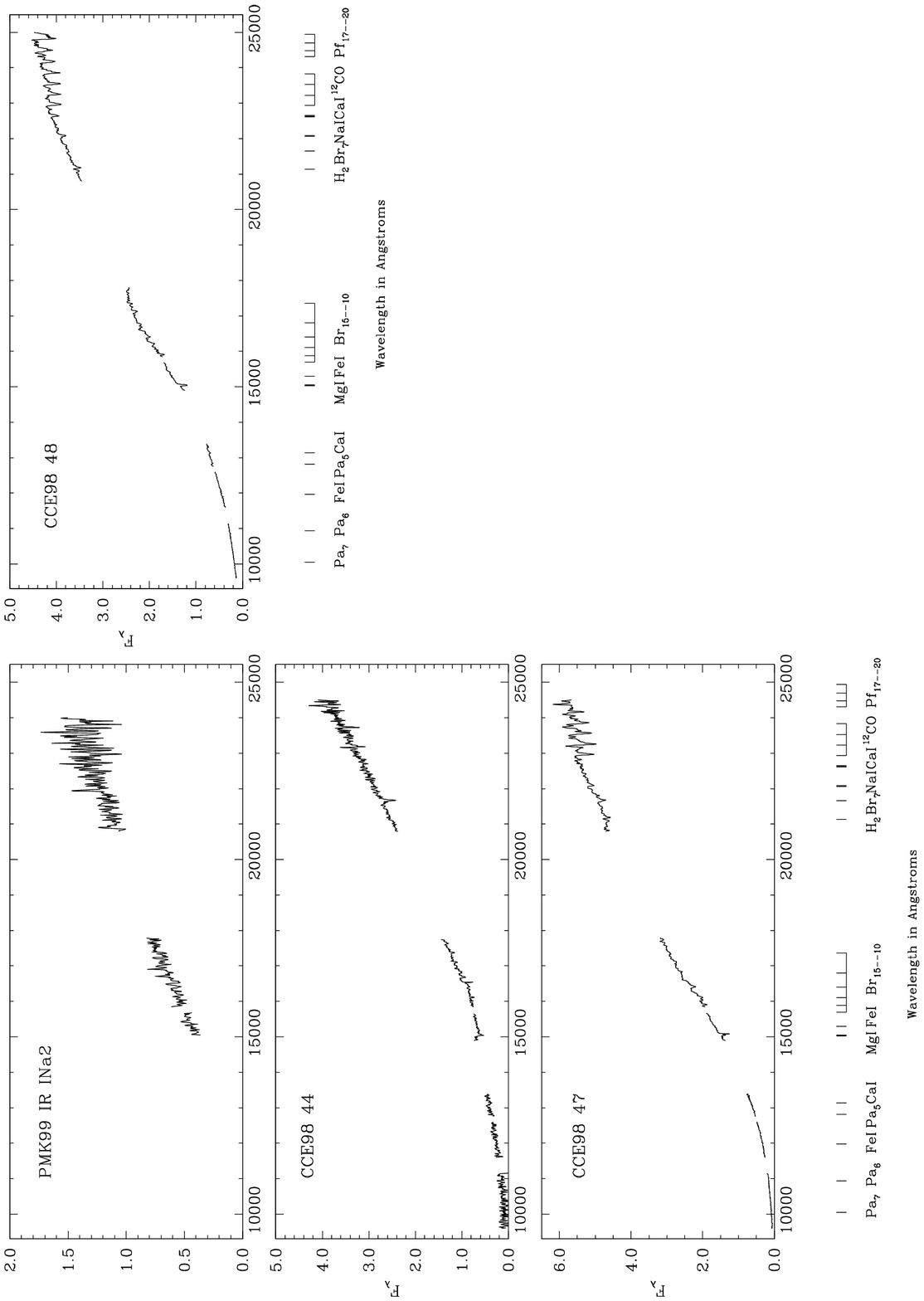}
\centering
\vskip -0.3in
\caption{$JHK$ band spectra of candidate young stellar objects in
the Chamaeleon I dark. Prominent atomic and molecular features in the
observed spectral range are labeled.}
\label{Fig2}
\end{figure}        

\begin{figure}
\centering
\includegraphics[width=7.0in]{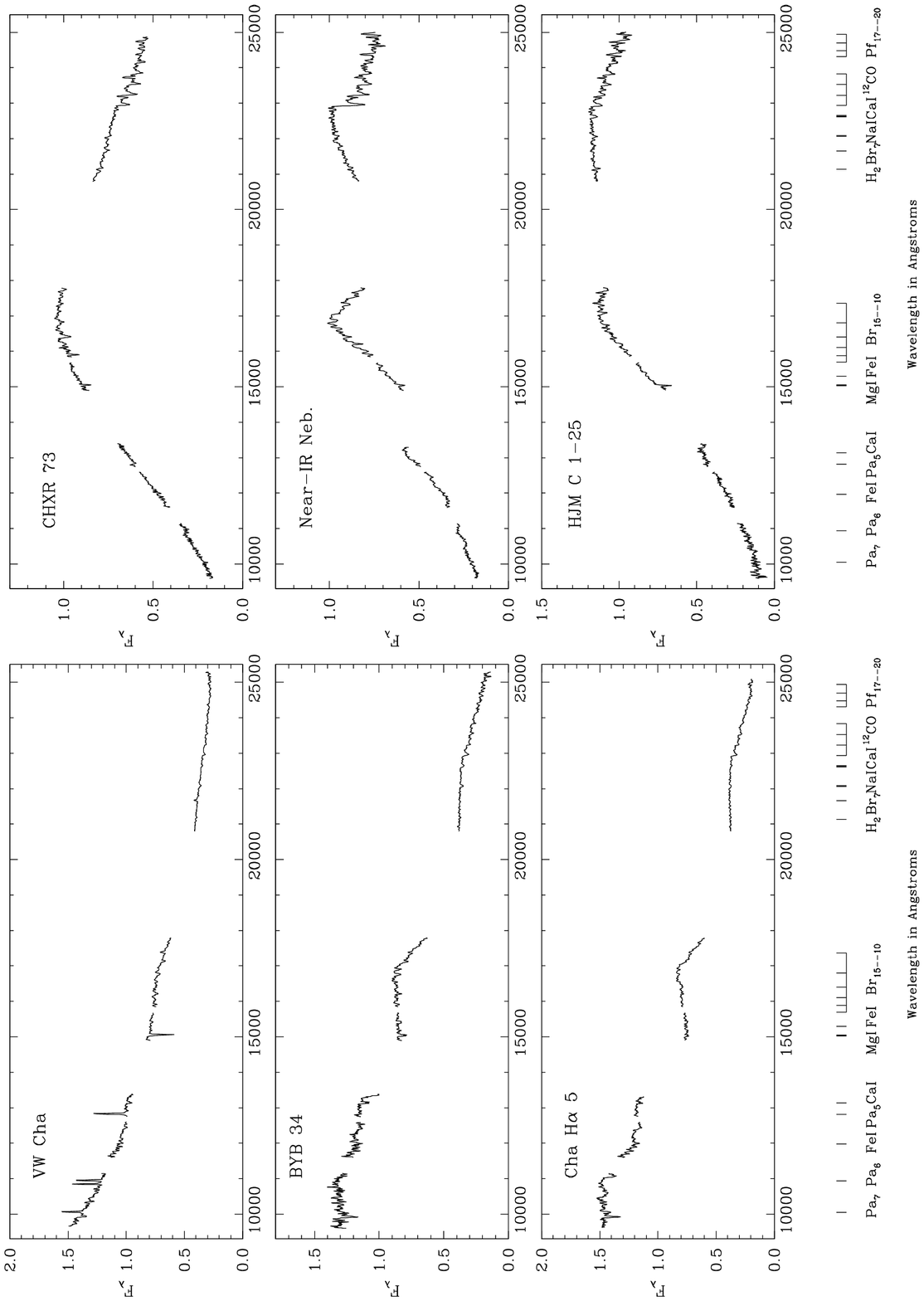}
\end{figure}   

\begin{figure}
\centering
\includegraphics[width=7.0in]{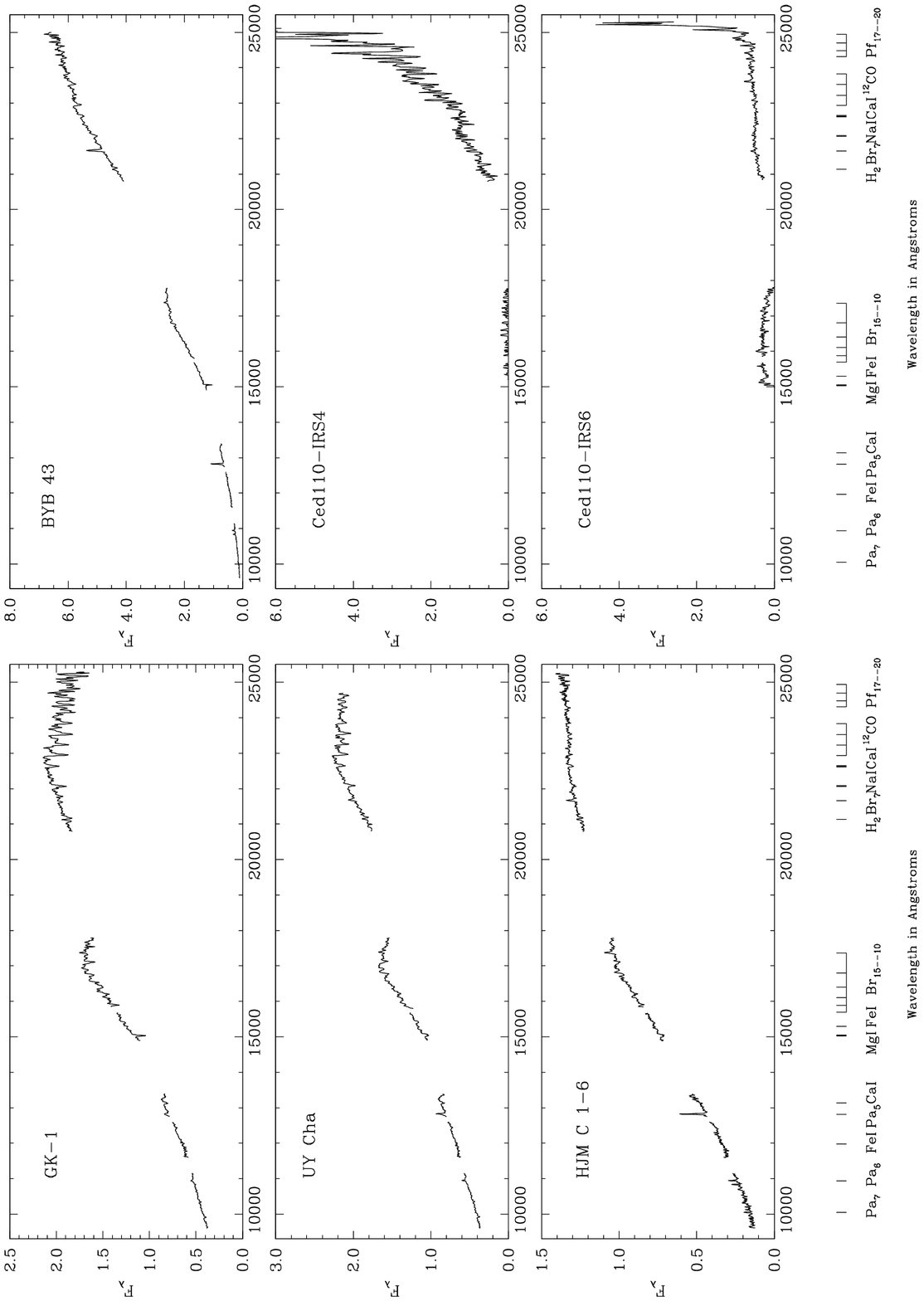}
\centering
\vskip -0.3in
\caption{$JHK$ band spectra of 12 of the previously known young stellar
objects for which we obtained near-infrared data.
Prominent atomic and molecular features in the
observed spectral range are labeled.}
\label{Fig3}
\end{figure}  

\citet{grla96} presented an atlas of low resolution spectra of
$\sim$ 100 young stellar objects in different star-forming
regions, including Taurus, $\rho$ Ophiuchi, R Cra, IC 5146, Scorpio, and Lupus. 
These authors describe near-infrared spectroscopic characteristics
of class I, II, and III objects as well as flat-spectrum and FU Orionis
objects. They found that, in general, the spectral shapes correlate
with the spectral energy distribution class and that equivalent widths
of absorption features decrease from the class III to the class I sources, as
a consequence of the increasing amount of circumstellar material.
In addition many young stellar objects (usually 
class II and I) have atomic hydrogen lines in emission. 

For the previously known young stellar objects observed we find 
a similar correlation between the spectral shape and the classical (class II)
or weak emission (class III) T Tauri type. In general, class II objects have rising or
flat spectral shapes. Class III stars have decreasing or flat spectral shapes. 
We find candidate sources have rising, flat, and decreasing shapes.
The first and second groups probably correspond to class I-II members of
the cloud and the rest to potential class III young stellar objects. 

Tables 5 and 6 list equivalent widths corresponding to the strongest lines
detected in the spectra of the observed targets. 
For the SOFI data we estimate an uncertainty of $\sim$ 1-2 \AA~ in the
Pa$\beta$, Br$\gamma$,  Na I doublet (2.206 and 2.209 $\mu$m), and Ca I triplet
(2.261, 2,263, and 2.266 $\mu$m) equivalent widths. The combined 
CO $\nu$ $=$ 2-0 and 3-1 bands have a slightly worse precision of 
$\sim$ 3 \AA. The OSIRIS spectra in general have poorer S/N ratio 
in relation to SOFI data. We estimate an error of 2-3 \AA~ in
our measurements of Pa$\beta$, Br$\gamma$,  Na I doublet, and Ca I triplet. 
The OSIRIS spectra only cover CO $\nu$ $=$ 2-0, and 3-1 bands.  We 
estimate an uncertainty of $\sim$ 4-5~ \AA~ in the two CO bands covered 
by this instrument. The three stars observed with both instruments
(VW Cha, SZ 119, and SZ 124) have equivalent width
measurements that agree within the estimated uncertainties. 
We notice the SOFI spectra cover the four CO $\nu$ $=$ 2-0, 3-1, 4-3, and 5-3
bands. In Tables 5 and 6 we provide the combine CO equivalent width
corresponding to the $\nu$ $=$ 2-0 and 3-1 bands to properly compare these
measurements with those obtained for the OSIRIS data. 

Stars in Tables 5 and 6 have, on average,
similar or slightly smaller equivalent widths than M type standards
\citep{klha86,wal00}, with exception of few sources (such as 
PMK99 IR Cha INa1, Ced 110 IRS 4, Ced 110 IRS 6, and Glass I) with no absorption lines
present, at the spectral resolution used. \citet{grla96} also noticed that
many young stellar objects in their sample (about 100 stars belonging to $\rho$
Oph, Taurus, and other star-forming regions) have atomic (Na I and Ca I) and
CO equivalent widths in the same range as late-type MK standards. 

Figure 4 shows the Ca I vs Na I equivalent widths for the groups
of previously known class II (CTTS) and class III (WTTS) objects observed.
For the Ca I line equivalent width we obtain an average
of 3.30 $\pm$ 1.39 and 4.88 $\pm$ 1.59 for the class II and class III,
respectively.  The Na I line equivalent width gives similar results: 3.48 $\pm$ 1.59 for the
CTTS and 3.87 $\pm$ 1.32 for the WTTS. Although the uncertainties are large, 
on average, the class II sources have slightly smaller equivalent widths compared
to the class III group.  Smaller equivalent widths may 
indicate the presence of circumstellar material partially filling-in these absorption
lines. A similar plot for the new sources gives only a marginal correlation between
the equivalent widths and the spectral shapes. We note that the poor S/N ratio in the spectra
of several candidate objects rendered uncertain the detection of spectral line features.
Better S/N data are required to confirm this initial correlation.

\begin{figure}
\centering
\includegraphics[width=17cm]{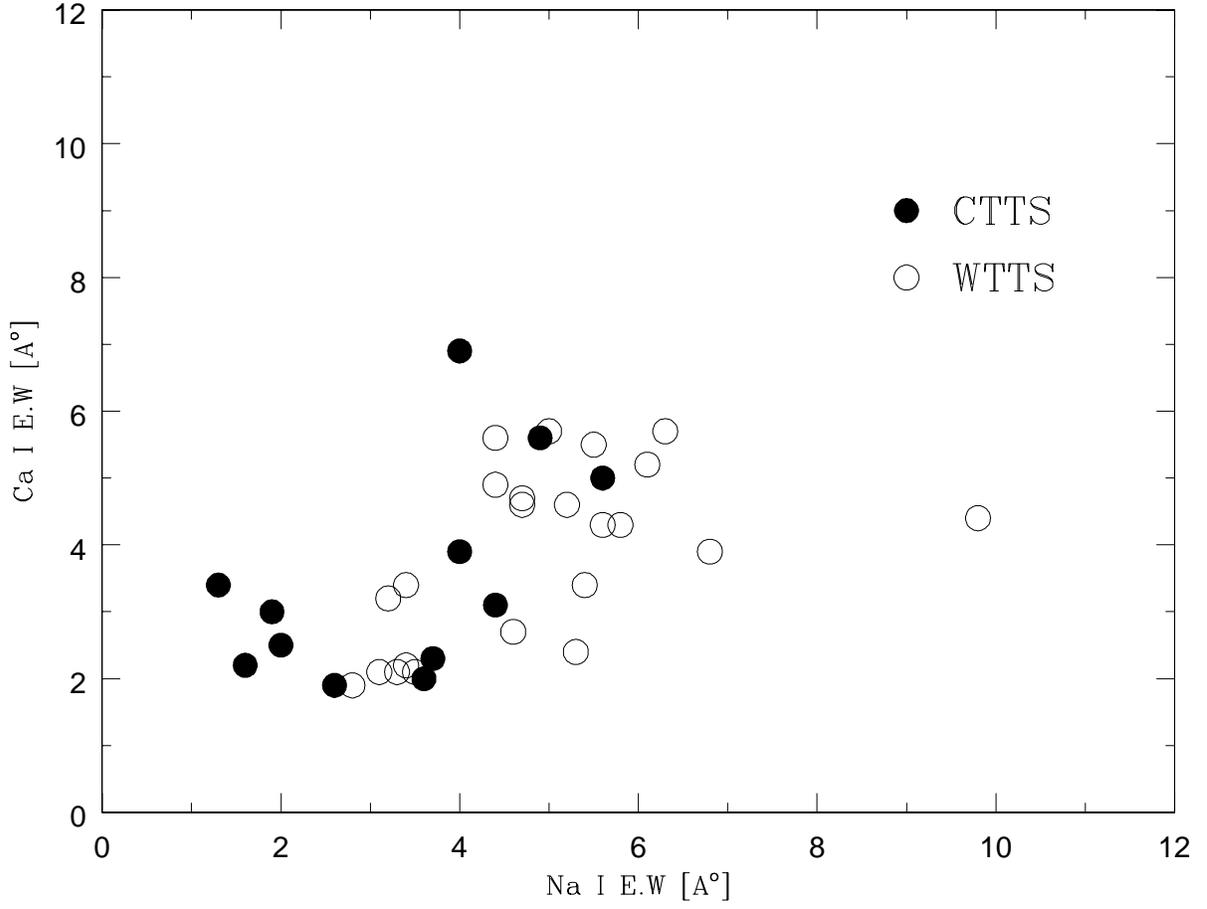}
\caption{Ca I vs Na I equivalent widths (in \AA) for previously known class II
(CTTS) and class III (WTTS) sources.}
\label{Fig4}
\end{figure}  

Several stars (such as CCE98 24, VW Cha,
HJC C 1-6, and BYB 43) also show H atomic lines in emission; the strongest are
Pa$\beta$ and Br$\gamma$. We notice an apparent tendency of a higher frequency
of H I emission lines among previously known young stellar objects than among the
new sources. Twenty of the 63 previously known young stars have Pa$\beta$
or Br$\gamma$ in emission, whereas only one of the 29 measured candidate young
stars shows these features.  However the modest S/N ratio of some of the new stars spectra
may explain at least in part this tendency. Higher S/N data are necessary 
to properly consider this issue.

\subsubsection{The PMK99 IR Cha INa1 Spectrum}

Figure 5 shows an enlarged spectrum of the source PMK99 IR Cha INa1,  
that has molecular hydrogen in emission. In particular H$_2$ 1-0 Q (1--8) as
well as H$_2$ 1-0 S (1--0) are clearly detected. 
PMK99 IR Cha INa1 is a class I source in the cloud identified by
\citet{per99} roughly located at the center of the high
velocity bipolar outflow found by \citet{mat89}.   
Based on the physical properties of PMK99 IR Cha INa1 (steep spectral energy
distribution and association with a small near-infrared nebula) 
\citet{per99} proposed this star rather than HM 23 (a class II source 
lying in the region) as the exciting source of the bipolar outflow. HM 23 had been
originally associated with the CO flow as it was the nearest young star known 
in the vicinity of the flow. Molecular hydrogen lines in the spectrum of
PMK99 IR Cha INa1 are probably originated in the shock regions where
bipolar outflow impacts on the surrounding molecular gas
\citep{bur89, eis00, ros00}.

\begin{figure}
\centering
\includegraphics[width=17cm]{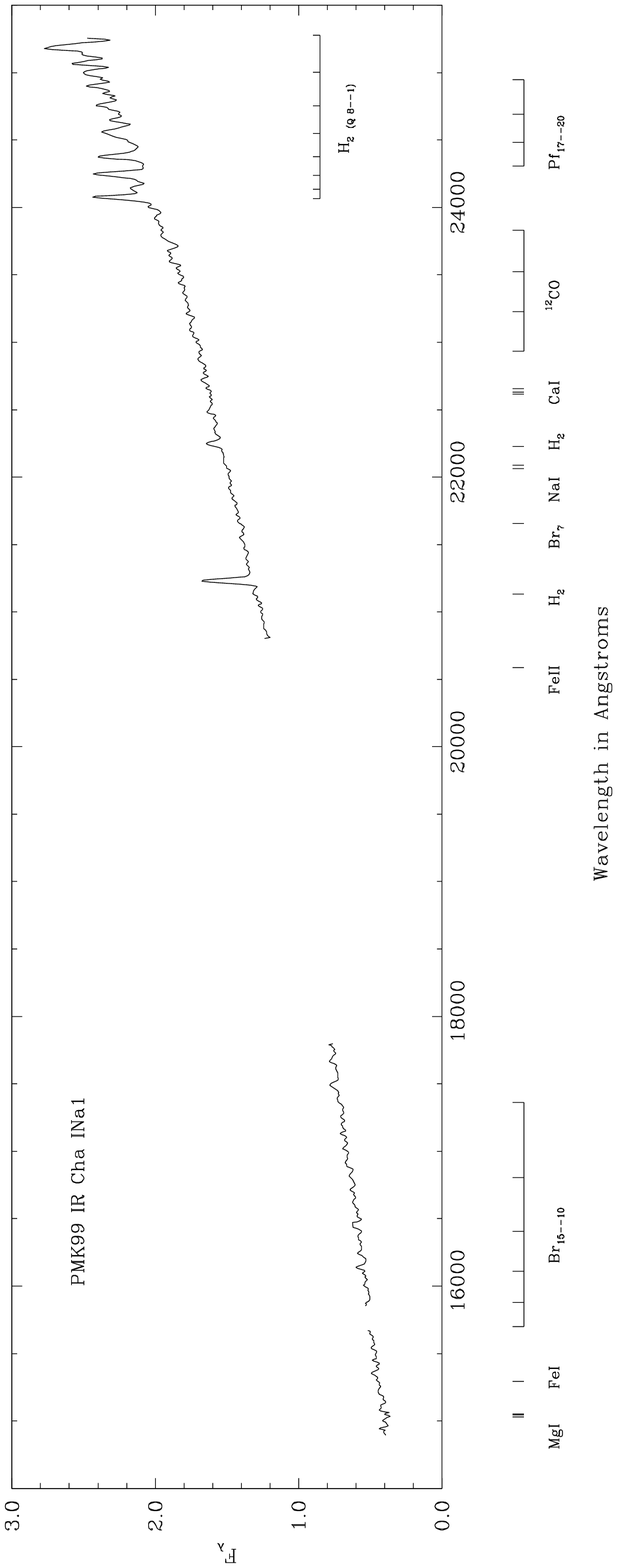}
\caption{$HK$ band spectrum of PMK99 IR Cha INa1.
Molecular hydrogen lines, \hbox{H$_2$ 1-0 Q (1--8)} and \hbox{H$_2$ 1-0 S
(1--0),} are in emission.}
\label{Fig5}
\end{figure}

\subsection{Spectral Type}

To derive spectral types for the candidate sources we applied the stellar water
vapor indexes proposed by \citet{wil99} and \citet{com00}. These indexes
are reddening-free and compare the intensity of the absorption bands near
2 and 1.9 $\mu$m, respectively. The strength of these bands are extremely sensitive
to the spectral types for M dwarfs.  \citet{wil99}'s
index $Q$ is defined by the average values of relative flux
density calculated in three narrow bands, F1 (2.07--2.13 $\mu$m),
F2(2.267--2.285 $\mu$m), and F3(2.40--2.50 $\mu$m), as
 
\begin{equation}
Q={\rm (F1/F2)(F3/F2)^{1.22}}.
\label{uno}
\end{equation} 

\noindent
This index can only be applied to the SOFI spectra. OSIRIS data have poor
S/N ratio at wavelengths longer than 23500 \AA. 

\noindent
The index ${\rm I_{H_2O}}$ \citep{com00} measures the relative fluxes in four narrow
band filters (0.05 $\mu$m width) centered at 1.675 $\mu$m, \hbox{1.750 $\mu$m,} 2.075
$\mu$m, and 2.25 $\mu$m. Denoting the fluxes at each narrow band filter by f1, f2, f3,
and f4, respectively, the index ${\rm I_{H_2O}}$ is expressed by:                                                                     
\begin{equation}
{\rm I_{H_2O} = (f1/f2)(f4/f3)^{0.76}}.
\label{dos}
\end{equation} 

\noindent
The ${\rm I_{H_2O}}$ index allows us to derive spectral types for our
candidate sources observed with both instruments as, in general, involves
spectral regions of good S/N ratios.

We notice that the $Q$ and ${\rm I_{H_2O}}$ indexes yield values which are
independent of the extinction.  They provide an appropriate tool to estimate spectral
types for young stellar objects usually heavily obscured by the parent cloud
dusty material. 

\citet{wil99} \citep[see also][]{cus00} obtained a linear relation
to determine the M sub-type as function of $Q$ using M dwarf standards with
optically known spectral types \citep[see equations (2) in][]{wil99,cus00}.
These relations have been derived from standards whose
spectra were telluric corrected using A0 V stars but whose
continuum shapes were not restored (i.e., without multiplying by the 
Planck function spectrum at the temperature of the corresponding standard). 
As explained in Section 2 we used
both G3-5 and O8 telluric standards for our SOFI data and thus our
atmospheric telluric divided spectra (without correction for the slope of
the standard spectrum) are not entirely comparable to those of \citet{wil99} 
and \citet{cus00}. Nevertheless, \cite{gope02} noticed little dependence on the
spectral type of the telluric standard in the analysis of ISOCAM sources of the
Chamaeleon I dark cloud. 

To check the applicability of 
these calibrations to our data we plot, in Figure 6, the optical spectral
types vs the spectral types derived from the index $Q$ for previously known objects.
In this case we show the plot corresponding to the spectra telluric corrected using
the O8 standards.  The results are basically the same when we use the G5 stars. 
Our spectra of SZ 23 and GK-1 have relatively poor S/N ratio in the 2.4 -- 2.5 $\mu$m region
and the fluxes in the band F3 are very uncertain. These two objects were
not included in Figure 6. We fit a linear relation to the data on this plot
with a correlation coefficient r $=$ 0.95. All spectral types derived from the index $Q$
agree within 1.5 sub-type with the optical types. This difference is comparable to the
accuracy expected by \citet{wil99} and \citet{cus00} for their calibrations. 

\begin{figure}
\centering
\includegraphics[width=17cm]{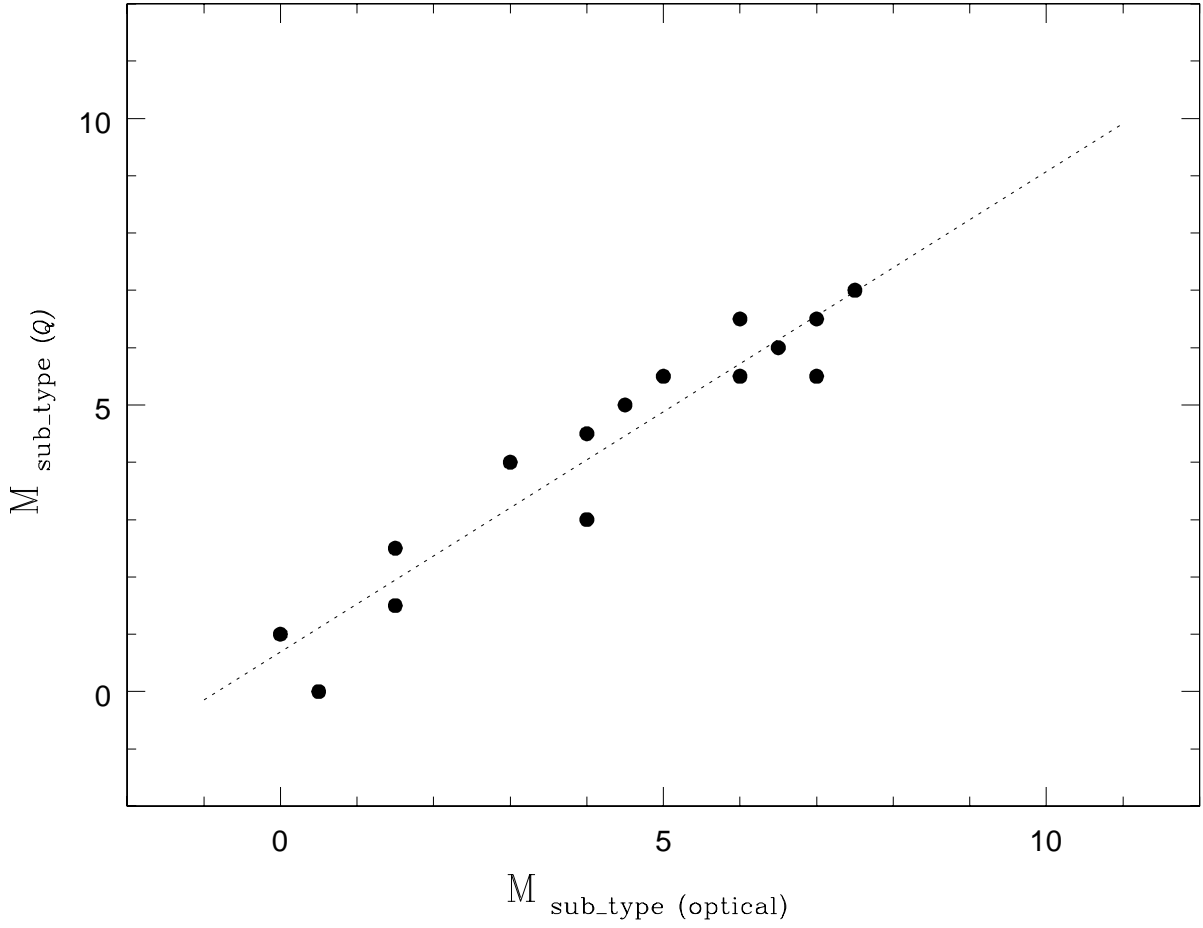}
\caption{Optical M{$_{\rm sub-type}$} vs $Q$-index-derived
M{$_{\rm sub-type}$} for previously known objects observed with SOFI.
The dash line indicates a linear fit to the data.}
\label{Fig6}
\end{figure}    

We derive the index ${\rm I_{H_2O}}$ for previously known objects
with optical spectral type observed with both instruments and obtain the
following linear least-squares fit to the data:
 
\begin{equation}
{\rm M_{sub\_type}} =  (-14.22 \pm 1.25) + (15.07 \pm 1.08) \times {\rm
I_{H_2O}},
\label{tres}
\end{equation}                        

\noindent
with a correlation coefficient r $=$ 0.94. We estimate a maximum uncertainty of
about 1.7 sub-classes in this calibration, considering {\it typical} errors in
our measurements of the index ${\rm I_{H_2O}}$ and the correlation coefficient
of equation \ref{tres}. \citet{gope02} derived a similar relation of
considerably lower precision based only on six previously known observed
objects. We also determined equation (3) using only SOFI and OSIRIS data
separately.  Figure 7 shows the optical spectral types vs the spectral
types derived from the index ${\rm I_{H_2O}}$ for previously known objects.
Both calibrations agree within the 1.7 sub-classes uncertainty.    
Poor telluric corrections rendered unreliable the fluxes in one or more bands
used by this index for several objects such as SZ 119, CHXR 78c, UY Cha,
HO Lup, and SZ 77 and thus these stars were omitted in the derivation of
equation \ref{tres}.

\begin{figure}
\centering
\includegraphics[width=17cm]{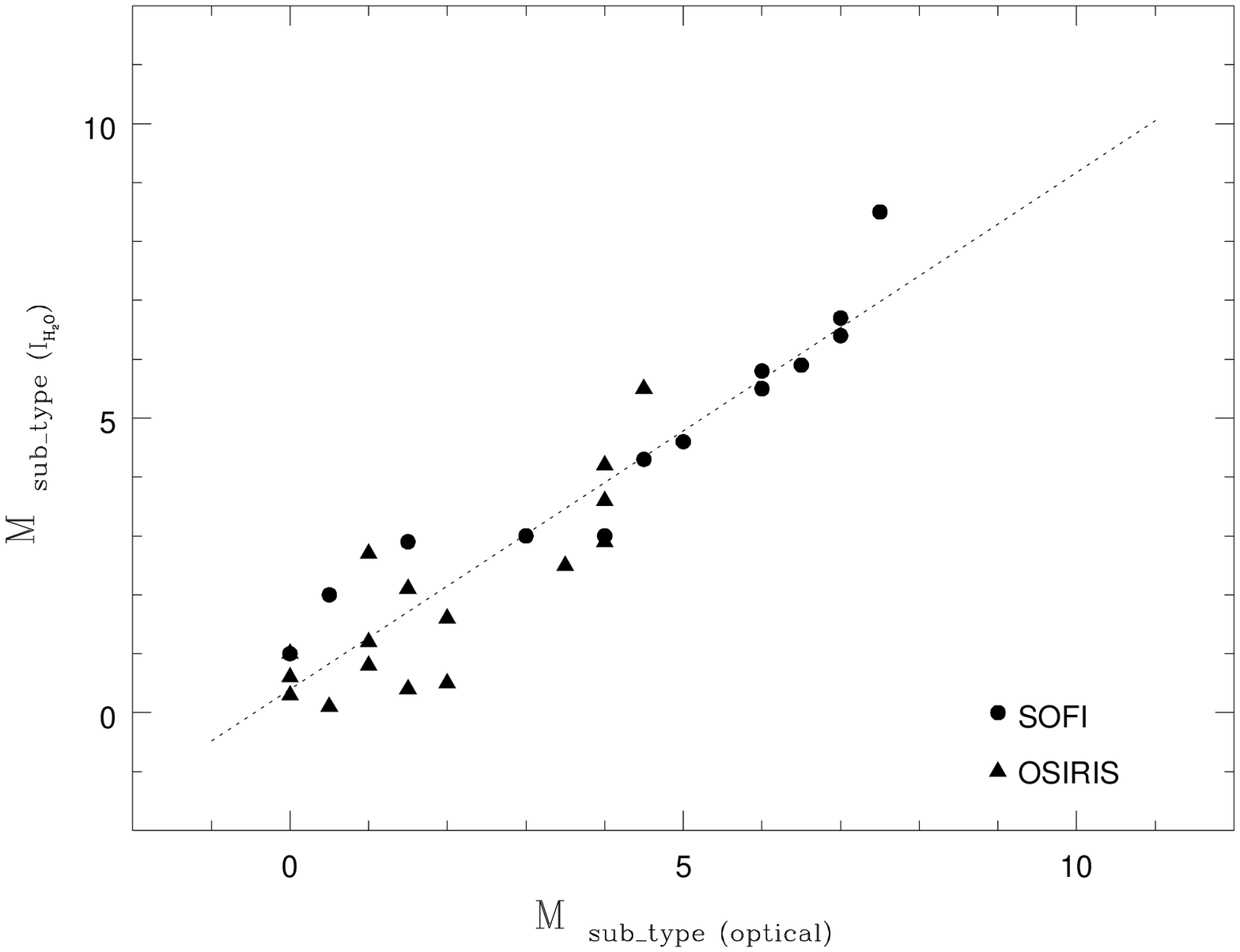}
\caption{Optical M{$_{\rm sub-type}$} vs ${\rm
I_{H_2O}}$-index-derived M{$_{\rm sub-type}$}
for previously known objects observed with SOFI and OSIRIS.
The dash line indicates a linear fit to the data.}
\label{Fig7}
\end{figure}        

In Table 7 we give spectral types derived from the application of these
indexes for our candidate sources. Several objects in our sample such as CCE98 20, 
CCE98 35, CCE98 39, and CCE98 46 have very poor S/N spectra and thus spectral
type derivations would be very uncertain.  Other two near-infrared selected objects
(CCE98 9 and GK2001 15) have better S/N spectra in relation to the ones
mentioned above but still rather poor to attempt a reliable spectral type
determination.  Finally two objects with relatively good S/N spectra
CCE98 40 and CCE98 44 have spectral types earlier than M0 justing
from the significant Br$\gamma$ equivalent width absorption lines (7 and 8 \AA, respectively) 
and the absence of Na I, Ca I and CO features in their spectra (see Table 5).

Table 8 lists a group of previously known objects without optical spectral
types in the literature for which
we applied the water vapor indexes. Poor telluric corrections around 
2.05 $\mu$m compromised the fluxes in the F3 and f3 bands ($Q$ and
${\rm I_{H_2O}}$ indexes, respectively) and prevented
us from obtaining spectral types for Ced 110 IRS 4 and Ced 110 IRS 6.  
HJM C 2-3 has a significant Br$\gamma$ equivalent width ($\sim$ 12 \AA) and
no additional features indicative of a late spectral type. This object is
probably earlier than M0. 

For objects with SOFI spectra available we derive spectral type from
both indexes and obtain a good agreement within 1.7 sub-class. However,
it was not always possible to accurately measure the flux in the F3 band 
corresponding to the index $Q$ due to poor S/N ratio in this spectral region. 
In these cases we adopted the spectral type given by the index
${\rm I_{H_2O}}$. 

To derive effective temperatures, intrinsic colors, and bolometric corrections
for our targets from the estimated spectral types 
we adopted the corresponding calibrations obtained by \citet{wil99}.
These relations comprise spectral types from M2 to M9.  
We supplemented these determinations with \citet{keha95} relations 
in order to include M0-M1 stars. 

\subsection{Effects of veiling and surface gravity}

The thermal emission from warm dust grains in circumstellar disks around young
stellar objects may affect the observed water vapor bands and thus the spectral type
estimates for our targets. This infrared excess emission can be characterized
by {$\rm r_\lambda = F_{\lambda}^{exc}/F_{\lambda}^{phot}$}, where {$\rm F_{\lambda}^{exc}$}
is the measured flux, including the excess (circumstellar) emission,
and {$\rm F_{\lambda}^{phot}$} is the flux due to the underlying stellar photosphere. 
\citet{mey97} analyzed the near-infrared properties of T Tauri stars located in
the Tarus-Auriga molecular cloud and derived median values of the J-, H-, and 
K-band excesses for the group of the CTTS and for the WTTS.  They obtained
$<{\rm r_J}>$ $=$ 0.0, $<{\rm r_H}>$ $=$ 0.2, and $<{\rm r_K}>$ $=$ 0.6
for the CTTS and $<{\rm r_J}>$ $=$ $<{\rm r_H}>$ $=$ $<{\rm r_K}>$ $=$ 0.0 for
the WTTS. In Tables 7, 8, and 9 we give r$\rm _K$, the $K$-band excess, estimated for our target
objects using the following relation,
                                                      
\begin{equation}
{\rm E(H-K) = (H-K) - (H-K)^{phot} - 0.077 \times A_V
       = -2.5 \times log [1/(1+r_K)]}
\label{cuatro}
\end{equation}          

\noindent
\citep[see][]{mey97}. In this expression {$\rm (H-K)$} is the observed
color (see Tables 3 and 4) and {$\rm (H-K)^{phot}$}
corresponds to the intrinsic or photospheric color. 
The third term in equation (\ref{cuatro}) is related to the excess in
color due only to the extinction to each individual target. To derive {$\rm A_V$} 
we applied the \citet{rile85} reddening law, {$\rm A_J = 0.28 A_V$}, with 

\begin{equation}
{\rm A_J = 2.63 \times E(J-H)}
\label{cinco}
\end{equation}                  

\noindent
where, {$\rm E(J-H) = (J-H) - (J-H)_o$}, is the color excess, and
{$\rm (J-H)_o$}, the intrinsic color.

Tables 7, 8, and 9 list $\rm A_J$ for the observed targets as well as
the corresponding values of r$_K$. 
This estimate provides only a lower limit to the veiling as equation
(\ref{cuatro}) assumes r$\rm _H$ $=$ r$\rm _J$ $=$ 0
\citep[see][]{mey97,wil99,cus00}. However we expect the K-band excess, {\rm r$_K$},
provides the main contribution to the total infrared excess, {$\rm r_\lambda$}, 
in view of the median values for the CTTS in Taurus derived by \citet{mey97}. 

\citet{wil99} have simulated the spectra expected from young stars
surrounded by circumstellar matter, adding the
disk contribution to the spectra of known standard stars.
These authors have then compared the index $Q$ calculated for stars
of known spectral types {\it with} and {\it without} the disk emission.
The spectral types obtained from star+disk systems that have veiling
effects of r$\rm _K \sim$ 0.2 are about 1 sub-class earlier than those
derived for objects without the disk contribution. This effect increases
as {\rm r$_K$ increases and can account for $\sim$ 2-3 sub-classes
earlier for r$\rm _K \sim$} 0.6.                                                                     

Most objects in Tables 7, 8, and 9 have r$\rm _K$ $<$ 0.2--0.3 with the
exception of PMK99 IR Cha INa1 (r\rm $_K$ $=$ 0.7).
We expect our spectral type determinations
using the water vapor indexes to be little affected by the effects of veiling except 
for PMK99 IR Cha INa1 that may actually have a spectral type
$\sim$ 3 sub-types later than that given in Tables 7 according to its
large K-band excess. It is also likely that in this case at least the 
H-band excess contribution be not negligible and then 
increases the total effect. 
In the Table 7 (in brackets) we adopt a minimum veiling correction of 3
sub-types later and indicate the corresponding derived parameters. We include
this correction on the estimated mass and age for this object
in section 3.6.  We notice that any additional veiling correction would
shift the object practically horizontally on the HR diagram corresponding
to decreasing values of mass and age. For example an additional 
increase of the spectral type in 0.5 sub-class (i.e., M7 spectral type)
would give a mass of 0.05 ${\rm M\sun}$ and an age of 2 $\times$ 10$^5$ yr
for this object. 

Pre-main sequence stars are known to have surface gravities
intermediate between those of dwarfs (luminosity class V) and giants
(luminosity class III) \citep[e.g.,][]{grme95, grla96}. However
we have determined spectral types for our targets neglecting this
effect, in particular when applying the $Q$ index versus the M spectral type relation
derived by \citet{wil99} and \citet{cus00}. These authors used M main sequence
standards to obtain this calibration. To derive equation (3), the ${\rm
I_{H_2O}}$ index calibration, we only used pre-main sequence stars with
optically known spectral types and thus this relation may be more appropriate, 
in terms of the gravity effect, to obtain spectral types 
for our candidate young stellar objects. 

Several authors \citep[see, for example,][]{wil99, cus00, gope02} have
analyzed this effect on spectral type determinations and in general conclude
that the surface gravity effect introduces a {\it typical}
uncertainty of about 1--2 sub-classes. This estimate is appropriate for
our water vapor index based spectral classification. 

\subsection{Luminosities, masses, and ages }

We used the $J$ magnitudes to estimate bolometric luminosities. 
This spectral band is less affected by contamination from circumstellar
infrared excess emission than $K$ or $H$. The bolometric
luminosities are calculated from the following expressions:

\begin{equation}
{\rm Log (L_{bol} / L_{\sun}) = 1.89 - 0.4 \times M_{bol}} 
\label{seis}
\end{equation}        

\begin{equation}
{\rm M_{bol} = m_J - A_J - DM + BC_J} 
\label{siete}
\end{equation}       

\noindent
where, DM $=$ 6.0 is the distance modulus and {$\rm {BC_J}$} 
is the $J$ band bolometric correction. {$\rm {A_J}$}\footnote{For a few objects
with no $H$ magnitude measurements we used the $K$ band data in combination with
the \citet{rile85} reddening law ($\rm {A_J = 1.66 \times E(J-K)}$, with
${\rm E(J-K) = (J-K) - (J-K)_o}$) to estimate the bolometric
luminosities.} and {$\rm m_J$} 
are the extinction and the apparent magnitude, respectively. 
Luminosities for the observed targets are given in Tables 7, 8, and 9. Our determinations agree within
a factor of $\sim$ 2 with previous derivations for the already known
objects in the Chamaeleon I and Lupus clouds \citep[cf.][]{law96, com00, kra91, hug94}. 

We chose pre-main sequence evolutionary tracks and isocrones from D'Antona \& Mazzitelli
(1994,1998)\footnote{Available at http://www.mporzio.astro.it/~dantona/prems.html.}
to derive masses and ages for the observed sources. This model
provides consistent and plausible results for our group of new targets, in reasonable
agreement with the higher mass members of the cloud.
In addition, D'Antona \& Mazzitelli's (1994,1998) calculations include stars in a wide
range of masses, from 3 ${\rm M\sun}$ to objects well below the H burning limit
(0.017 ${\rm M\sun}$) and thus allows us an uniform and consistent derivation of
masses and ages for our targets. Tables 7, 8, and 9 list these parameters. 
Figure 8 shows the location of new and previously known Chamaeleon I stars
spectroscopically observed on the HR diagram.

\begin{figure}
\centering
\includegraphics[width=17cm]{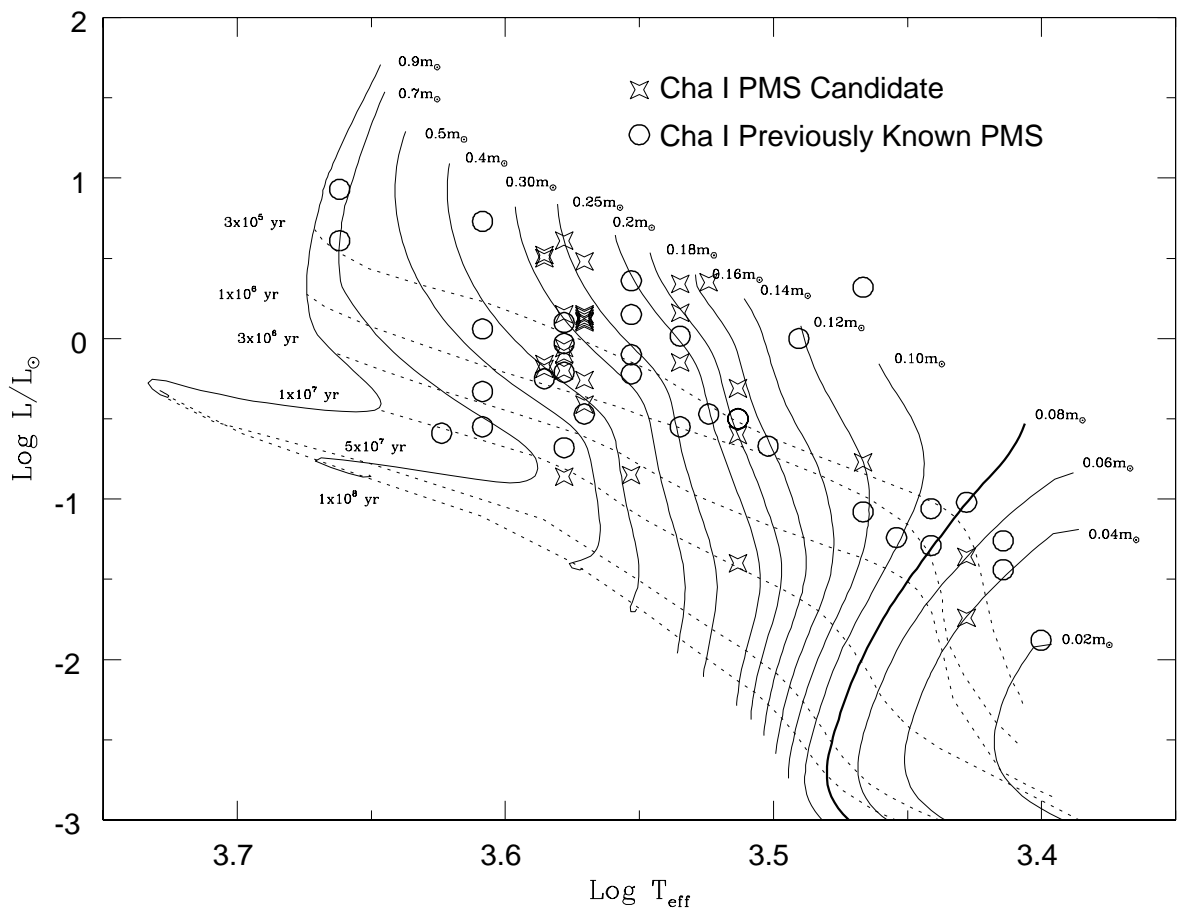}
\caption{HR diagram showing the positions of the candidate
and previously known pre-main sequence
stars spectroscopically observed in the Chamaeleon I dark cloud.
Pre-main sequence evolutionary tracks, indicated
with continuous lines, are from D'Antona \& Mazzitelli (1998). The thick
continuous line corresponds to 0.08 ${\rm M\sun}$, the H burning limit. The
dashed lines correspond to the isocrones calculated also by the same
authors.} 
\label{Fig8}
\end{figure}   

We derived a mass of 0.19 ${\rm M\sun}$ and an age of 4 $\times$ 10$^6$ yr
for the source PMK99 IR Cha INa1 using the spectral type M3.5 given in Table 7.
However as discussed in section 3.5 this object may actually have
a spectral at least type 3 subtype later due to the high veiling (r$\rm _K$ $=$ 0.7).
To correct for the veiling effect we adopt a M6.5 spectral type and
recalculate the corresponding bolometric luminosity
($\rm Log (L_{bol}/L_{\sun}) = -$1.36).
This shifts the star to the right practically horizontally on
the HR diagram, corresponding to decreasing values of mass and age
(0.06 ${\rm M\sun}$ and 4 $\times$ 10$^5$ yr, respectively).
In Table 7 we indicate in brackets the veiling corrected parameters. 
To our knowledge PMK99 IR Cha INa1 is the first 
substellar object associated with a bipolar outflow.

The candidate Chamaeleon I young stars (Figure 8) have masses
between $\sim$ 0.7 and 0.04 ${\rm M\sun}$. The previously known members of the
cloud in Figure 8, roughly show the same range of masses and
with the exception of LkH$\alpha$ 332-17 ($\sim$ 2.4 ${\rm M\sun}$), 
one of the highest mass members of the cloud \citep[see also][]{law96}.  
GK2001 18 with a mass of 0.04 ${\rm M\sun}$ and PMK99 IR Cha INa1 with a mass
of 0.06 ${\rm M\sun}$ are the only two substellar objects detected among
the observed Chamaeleon I candidates. The new objects have ages between
$>$ 1 $\times$ 10$^5$ and $\sim$ 2 $\times$ 10$^7$ yr, roughly in agreement with 
the known members of the cloud \citep{law96}. 

Our mass determinations for the known members agree within a factor
of $\sim$ 2 with those derived by \citet{law96} and \citet{com00}. We
find that the age determinations agree within a {\it typical} factor of 4-5,
except for HM 15, HM 16, Glass I and Cha H$\alpha$ 1 -- 6, for which we
find a larger (about a factor of 10) difference in age. In the case
Cha H$\alpha$ 1 -- 6 this difference is a least in part due to 
the use of different isocrones. \cite{com00} based their
determination of ages on \cite{bur97} and \cite{bar98} models
whether we applied D'Antona \& Mazzitelli's isocrones. 

The Lupus previously known
young stars span a range of ages of $\sim$ 3 $\times$ 10$^5$ and
1 $\times$ 10$^6$ yr and masses between $\sim$ 0.5 and 0.14 ${\rm M\sun}$. 
The only exception is SZ 68, with a mass of 1.5 ${\rm M\sun}$.
Our determinations of mass and age agree within a factor of roughly 2 and 4, 
respectively with those derived by \citet{hug94}.  

\subsection{Mass and age distributions} 

\citet{law96} derived the mass and age distributions for the Chamaeleon I 
dark cloud based on $\sim$ 80 previously
known members of the cloud, adopting D'Antona \& Mazzitelli's (1994)
and \citet{swe94} models. The mass distribution rises from $\sim$ 2 ${\rm
M\sun}$, peaks at 0.5-0.6 ${\rm M\sun}$ and then falls towards lower masses.
They obtained a median stellar mass of 0.55 ${\rm M\sun}$ as representative
of their sample. The age distribution peaks at ages $<$ 5 $\times$ 10$^6$ yr
indicating a roughly constant rate of star formation during the last 5
$\times$ 10$^6$ yr and a rapidly declines towards older ages.            

We combined our targets with those from \cite{gast92}, 
\cite{law96}, \cite{com00} and \cite{gope02} to
determine the mass and ages distribution for 145 objects belonging to the Chamaeleon I dark cloud.
To analyze an homogeneous as possible data sample we estimated bolometric
luminosities and effective temperatures for all the cloud
members in the same manner as for the objects in Tables 7 and 8 (see Sections 3.4 and 3.5).
We then used D'Antona \& Mazzitelli's (1994,1998) evolutionary tracks and isocrones to
derive masses and ages. Figure 9 indicates the mass and age distributions.  
The median mass is 0.30 ${\rm M\sun}$. 
For the age distribution we obtained a median of 5 $\times$ 10$^5$ yr.

\begin{figure}
\centering
\includegraphics[width=17cm]{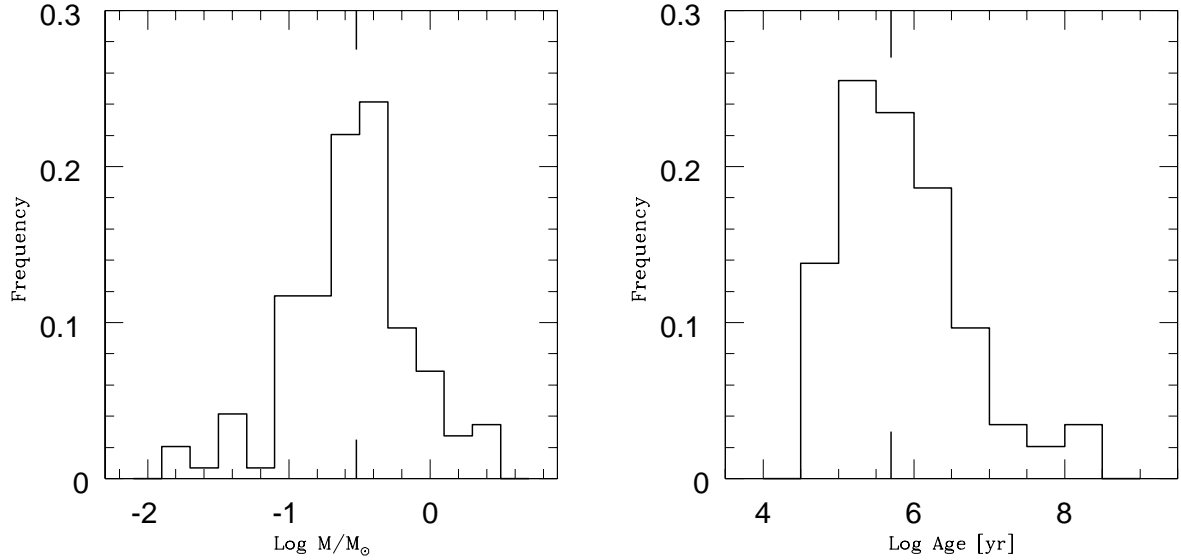}
\caption{Mass and age distributions for the Chamaeleon I stars
using the D'Antona \& Mazzitelli evolutionary tracks and isocrones.
The medians of each distributions are marked at the upper and lower axis. For
the mass distribution we derive a median of 0.30 ${\rm M\sun}$. The median age is
\hbox{5 $\times$ 10$^5$ yr.}}
\label{Fig9}
\end{figure}     

The mass histogram rises from about 2.5 ${\rm M\sun}$ down to 0.4 ${\rm
M\sun}$, then falls off. We note a substantial decrease in the median mass of the stars in the
Chamaeleon I dark cloud, from $\sim$ 0.55 ${\rm M\sun}$
\citep{law96} to about $\sim$ 0.30 ${\rm M\sun}$. This is mainly due to
the increasing number of mostly low mass and fainter members of the cloud
detected by recent deep infrared survey of the cloud.
The distribution of ages indicates an active star-formation episode within
the last $\sim$ 5 $\times$ 10$^5$ yr and a decreasing rate at older ages (few $\times$ 
10$^7$ yr). 

\citet{com00} intensively surveyed the central 300 arcmin$^2$
of the Chamaeleon I dark cloud, at different wavelengths combining
both photometric and spectroscopic information. They detected 13 new
members of the cloud and used \citet{bur97} and \cite{bar98} models to
identify 4 bona fide brown dwarfs, 6 transition
(stellar/substellar) objects and 3 very low mass objects.
In total this area contains 22 young stars with masses $<$ 1 ${\rm M\sun}$.
Assuming the pre-main sequence population in this region is essentially
complete they measured the IMF and obtained a roughly flat
behavior in logarithmic mass units within $\sim$ 1 
and $\sim$ 0.03 ${\rm M\sun}$. They also found that the majority
of identified members in the region (19 out of 22) have ages
near 2 $\times$ 10$^6$ yr and only 3 objects are significantly older,
with ages of $\sim$ 2 $\times$ 10$^7$ yr.         

The IMF of different star-forming regions (such as $\rho$ Oph, Trapezium, 
IC 348, L1495E and Taurus) is approximately flat from the substellar
regimen to $\sim$ 0.6--0.4 ${\rm M\sun}$
\citep[see][and the references therein]{luh00}. We interpret
the fall off of the current IMF for the whole Chamaeleon I dark cloud as 
population selection effect. Assuming that the {\it true} shape of
the function is roughly flat in logarithmic mass bins from the
subsolar to the substellar regimen and that the census of young stars in the
cloud for masses $>$ 0.4 ${\rm M\sun}$ is basically complete, we estimate that
$\sim$ 100 stars with masses between 0.4 and 0.04 ${\rm M\sun}$
remain to be identified in the cloud.

\section{Summary and Conclusions} 

We present near-infrared low resolution (R $\sim$ 500) spectra
of 46 candidate young stellar objects in the Chamaeleon I cloud 
with {\it typically} $K < 12$. We also observed
63 previously known young stellar objects belonging to this
star-forming region and other southern hemisphere clouds. 
Both groups have similar spectroscopic characteristics. 
We used the water vapor indexes to derive
spectral types for the new sources and adopted D'Antona \& Mazzitelli (1998)
evolutionary tracks and isocrones to estimate masses and ages. 
The majority of the new objects have masses between 0.7 and
0.15 ${\rm M\sun}$; with only two sources with masses in the substellar regimen.
One of these stars (PMK99 IR Cha INa1, with a mass of $\sim$ 0.06 ${\rm M\sun}$)
is associated with the bipolar outflow detected by \citet{mat89} in the
northern part of the cloud. The other object, GK2001 18, has an estimated
mass of 0.04 ${\rm M\sun}$. 

The mass distribution rises from about 2.5 ${\rm M\sun}$, has a peak around 0.4 ${\rm
M\sun}$ and then falls towards lower masses. The median mass for the 145 compiled stars in the Chamaeleon I
dark cloud is $\sim$ 0.30 ${\rm M\sun}$, representing a significant
decrease with respect to the $\sim$ 0.55 ${\rm M\sun}$ derived by
\cite{law96} from $\sim$ 80 known members.  The scarcity of very low mass objects is
interpreted as an observational selection effect toward the fainter
(and thus least massive) members of the cloud.
We assume the {\it true} IMF of the whole cloud is roughly flat in logarithmic mass bins in the
interval 0.4 -- 0.04 ${\rm M\sun}$ as found by \citet{com00}
in the central 300 arcmin$^2$ region. If the current pre-main sequence population with
masses $>$ 0.4 ${\rm M\sun}$ is complete, we estimate that 
$\sim$ 100 stars with masses within these limits remain to be
identified in the cloud. The age distribution indicates an active star-formation episode within
the last $\sim$ 5 $\times$ 10$^5$ yr and a decreasing rate at older
ages (few $\times$ 10$^7$ yr).

The new 46 candidates analyzed in this paper are, on average, 
roughly 1.5 mag fainter in the $K$ band than those included in \cite{law96}
sample. These new objects include only two substellar objects;  
the rest of the stars have sub-solar masses down to $\sim$ 0.15  ${\rm M\sun}$. 
However, $\sim$ 150 fainter candidate members of the Chamaeleon I cloud ({\it typically}
$K >$ 13) remain to be spectroscopically characterized
\citep{cam98, per99, oas99, per00, goke01, per01}. 
These objects may contain at least part of the missing
population within the interval 0.4 --- 0.04 ${\rm M\sun}$.
Near-infrared spectra, in combination with pre-main sequence evolutionary tracks and
isocrones, would be useful to determine
the masses and ages of these objects. This information would
provide a precise determination the behavior of the IMF, particular
towards and into the substellar regimen for the whole cloud.

\acknowledgments

We thank Leonardo Vanzi (ESO) and Robert Blum (CTIO) for
technical support with SOFI and OSIRIS, respectively.
An anonymous referee provided useful suggestions that
improved the content and presentation of this paper.                
This research has made use of the SIMBAD database,
operated at CDS, Strasbourg, France. DM gratefully acknowledges support
from Universidad de Chile FONDAP 15010003 and DID project I021-98/2.

\clearpage


\clearpage 

%% If you are not including electonic art with your submission, you may
%% mark up your captions using the \figcaption command. See the 
%% No more than seven \figcaption commands are allowed per page, 
%% so if you have more than seven captions, insert a \clearpage 

%% Tables should be submitted one per page, so put a \clearpage before

%% Two options are available to the author for producing tables:  the
%% deluxetable environment provided by the AASTeX package or the LaTeX
%% table environment.  Use of deluxetable is preferred.
%%

%% Three table samples follow, two marked up in the deluxetable environment,
%% one marked up as a LaTeX table.

%% In this first example, note that the \tabletypesize{}
%% command has been used to reduce the font size of the table.
%% Note also that the \label command needs to be placed 

\clearpage

%% Text for table notes should follow after the \enddata but before
%% the \end{deluxetable}. Make sure there is at least one \tablenotemark
%% in the table for each \tablenotetext.

%% If you use the table environment, please indicate horizontal rules using
%% Do not put multiple tabular environments within a single table.
%% The optional \label should appear inside the \caption command.

\clearpage

%% Any table notes must follow the \end{tabular} command.

%% If the table is more than one page long, the width of the table can vary
%% from page to page when the default \tablewidth is used, as below.  The
%% individual table widths for each page will be written to the log file; a
%% maximum tablewidth for the table can be computed from these values.
%% The \tablewidth argument can then be reset and the file reprocessed, so
%% that the table is of uniform width throughout. Try getting the widths
%% from the log file and changing the \tablewidth parameter to see how
%% adjusting this value affects table formatting.

%% In this example, we have used the optional * argument to \\ to
%% instruct LaTeX to keep rows together on the same page. (See the
%% lines following the \cutinhead.) Using \\* to group together table
%% rows on the same page affects how the table breaks. Try taking
%% the *'s out and LaTeXing again to see the difference.

%% You can append references to a table using the \tablerefs command.

%% Tables may also be prepared as separate files. See the accompanying
%% sample file table.tex for an example of an external table file.
%% To include an external file in your main document, use the \input
%% command. Uncomment the line below to include table.tex in this
%% sample file. (Note that you will need to comment out the \documentclass,
%% \begin{document}, and \end{document} commands from table.tex if you want
%% to include it in this document.)


\begin{deluxetable}{llllll} 
\tabletypesize{\scriptsize}
\tablecolumns{6} 
\tablewidth{16cm}  
\tablecaption{Candidate Chamaeleon I young stellar objects} 
\tablehead{
\colhead{Name} & \colhead{$\alpha$(2000.0)}   & \colhead{$\delta$(2000.0)}    & \colhead{Instrument} & 
\colhead{Date}    & \colhead{Other ID.}}
\startdata 

CCE98 1 &             10 58 22.9 & $-$78 29 49 &     OSIRIS & 1999 May 08  & \\
CCE98 2 &             10 59 12.3 & $-$78 26 40 &     OSIRIS & 1999 May 08  & \\
CCE98 5 &             11 01 32.1 & $-$77 42 10 &     OSIRIS & 1999 May 07  & \\ 
CCE98 8 &             11 02 47.1 & $-$77 38 10 &     OSIRIS & 1999 May 08  & \\ 
CCE98 9 &             11 03 11.5 & $-$77 36 36 &     OSIRIS & 1999 May 08  & ISO-ChaI 16\\                
CCE98 12 &            11 04 11.2 & $-$77 50 13 &     SOFI   & 1999 Apr 12  & ISO-ChaI 40\\ 
GK2001 3 &            11 04 49.2 & $-$78 04 46 &     OSIRIS & 2002 Feb 28  & \\ 
CCE98 14 &            11 05 19.2 & $-$77 57 39 &     SOFI   & 1999 Apr 12  & \\
CCE98 16 &            11 05 24.1 & $-$76 07 44 &     OSIRIS & 1999 May 07  & \\
CCE98 17 &            11 05 54.4 & $-$77 38 43 &     SOFI   & 1999 Apr 10  & ISO-ChaI 69\\
CCE98 18 &            11 05 55.2 & $-$77 35 13 &     SOFI   & 1999 Apr 12  & ISO-ChaI 70\\
CCE98 19 &            11 06 15.9 & $-$77 37 51 &     SOFI   & 1999 Apr 10/12  & ISO-ChaI 76\\
CCE98 20 &            11 06 18.9 & $-$77 35 18 &     SOFI   & 1999 Apr 12  & \\
GK2001 15 &           11 07 06.5 & $-$76 37 17 &     OSIRIS & 2002 Feb 28  & \\
CCE98 21 &            11 07 09.6 & $-$77 18 47 &     SOFI   & 1999 Apr 12  & ISO-ChaI 91\\
CCE98 23 &            11 07 16.5 & $-$77 23 08 &     SOFI   & 1999 Apr 10/11 & ISO-ChaI 97 \\
CCE98 24 &            11 07 21.7 & $-$77 22 12 &     SOFI   & 1999 Apr 10/11 & BYB 35,ISO-ChaI 101\\
CCE98 25 &            11 07 23.7 & $-$77 41 25 &     OSIRIS & 1999 May 07  & ISO-ChaI 102\\
CCE98 26 &            11 07 36.9 & $-$77 35 19 &     SOFI   & 1999 Apr 11  & ISO-ChaI 107 \\
CCE98 27 &            11 07 37.5 & $-$77 33 09 &     SOFI   & 1999 Apr 11  & ISO-ChaI 109 \\
GK2001 18 &               11 07 46.9 & $-$76 15 17 &     OSIRIS & 2002 Feb 28  & \\
CCE98 32 &            11 08 04.2 & $-$77 38 43 &     OSIRIS & 1999 May 07  & ISO-ChaI 126\\
CCE98 33 &            11 08 12.1 & $-$77 18 54 &     SOFI   & 1999 Apr 11  & ISO-ChaI 130 \\
CCE98 34 &            11 08 12.9 & $-$77 19 13 &     SOFI   & 1999 Apr 11  & ISO-ChaI 131 \\
GK2001 21 &               11 08 22.6 & $-$76 49 19 &     OSIRIS & 2002 Feb 28  & \\
GK2001 24 &               11 08 44.5 & $-$76 13 29 &     OSIRIS & 2002 Feb 28  & \\
CCE98 35 &            11 08 57.2 & $-$77 43 28 &     SOFI   & 1999 Apr 10  & \\
ISO-ChaI 177 &        11 09 08.5 & $-$76 49 13 &     OSIRIS & 2002 Feb 27  & \\
CCE98 36 &            11 09 11.4 & $-$76 32 50 &     SOFI   & 1999 Apr 12  & OTS99 2 \\
CCE98 37 &            11 09 12.0 & $-$77 39 06 &     OSIRIS & 1999 May 07  & ISO-ChaI 179\\
CCE98 39 &            11 09 23.3 & $-$76 31 14 &     SOFI   & 1999 Apr 12  & \\
CCE98 40 &            11 09 26.9 & $-$76 33 33 &     SOFI   & 1999 Apr 12 & OTS99 13,HJM C 1-3 \\
PMK99 IR Cha INa1 &   11 09 29.4 & $-$76 33 28 &     SOFI   & 1999 Apr 10/12 & ISO-ChaI 192,CCE98 41,OTS99 15\\ 
PMK99 IR Cha INa2 &     11 09 36.6 & $-$76 33 39 &     SOFI   & 1999 Apr 10  & OTS99 18\\
CCE98 42 &            11 09 38.0 & $-$77 10 41 &     OSIRIS & 1999 May 07  & ISO-ChaI 196\\
CCE98 43 &            11 09 43.5 & $-$76 33 31 &     SOFI   & 1999 Apr 10  & HJM C 1-21,OTS99 21\\
CCE98 44 &            11 09 47.7 & $-$76 34 06 &     SOFI   & 1999 Apr 10  & HJM C 1-22,OTS99 25 \\
CCE98 46 &            11 09 53.4 & $-$77 17 16 &     SOFI   & 1999 Apr 11  & \\
CCE98 47 &            11 09 56.7 & $-$77 18 26 &     SOFI   & 1999 Apr 11  & ISO-ChaI 227 \\
PMK99 IR Cha INa3 &   11 10 01.5 & $-$76 32 50 &     SOFI   & 1999 Apr 10  & OTS99 30\\
PMK99 IR Cha INa4 &   11 10 04.5 & $-$76 33 09 &     SOFI   & 1999 Apr 10  & OTS99 32\\
CCE98 48 &            11 10 11.9 & $-$76 35 31 &     SOFI   & 1999 Apr 11  & ISO-ChaI 237,HJM C 1-8,OTS99 45 \\
CCE98 49 &            11 10 54.0 & $-$77 25 03 &     SOFI   & 1999 Apr 11  & ISO-ChaI 256\\
GK2001 42 &               11 11 21.1 & $-$78 05 20 &     OSIRIS & 2002 Feb 28  & \\
CCE98 50 &            11 11 29.5 & $-$76 09 29 &     OSIRIS & 1999 May 07  & \\
CCE98 51 &            11 11 32.4 & $-$77 28 12 &     OSIRIS & 1999 May 07  & \\     
\enddata

\tablerefs{ 
BYB: \cite{bau84}, CCE98: \cite{cam98}, GK2001: \cite{goke01},
HJM C: \cite{hyl82}, OTS99: \cite{oas99}, ISOCAM Cha INa: \cite{per99}, ISO-ChaI: \cite{per00}.}
\end{deluxetable} 

\begin{deluxetable}{lllll}
\tabletypesize{\scriptsize}  
\tablecolumns{5}
\tablewidth{13cm}
\tablecaption{Previously known young stellar objects observed}
\tablehead{
\colhead{Name} & \colhead{$\alpha$(2000.0)}   & \colhead{$\delta$(2000.0)}    &
\colhead{Instrument} & \colhead{Date}}   
\startdata                    
&   &  &   &   \\
\multicolumn{1}{l}{Chamaeleon I} \\   
&   &  &   &   \\
Hn 1    &             11 02 33.1 & $-$77 29 26 &     OSIRIS & 1999 May 08   \\ 
BYB 18  &             11 04 43.0 & $-$77 41 57 &     OSIRIS & 1999 May 06   \\
CHXR 73 &             11 06 26.5 & $-$77 37 38 &     SOFI   & 1999 April 10 \\
Ced 110 IRS 4 &       11 06 48.1 & $-$77 22 29 &     SOFI   & 1999 Apr 11   \\
CHXR 74 &             11 06 54.0 & $-$77 42 09 &     SOFI   & 1999 Apr 10   \\
UY Cha  &             11 06 59.5 & $-$77 18 54 &     SOFI   & 1999 Apr 12   \\
Ced 110 IRS 6 &       11 07 10.5 & $-$77 23 10 &     SOFI   & 1999 Apr 11   \\
Hn 6  &               11 07 13.0 & $-$77 46 39 &     OSIRIS & 1999 May 08   \\
Cha H$\alpha$ 1 &     11 07 16   & $-$77 35 54 &     SOFI   & 1999 Apr 10   \\
LkH$\alpha$ 332-17 &  11 07 20.7 & $-$77 38 07 &     OSIRIS & 1999 May 06   \\
BYB 34 &              11 07 35.3 & $-$77 34 05 &     SOFI   & 1999 Apr 10   \\
CHXR 26 &             11 07 37.0 & $-$77 33 33 &     OSIRIS & 1999 May 06   \\
Cha H$\alpha$ 2 &            11 07 42.7 & $-$77 34 00 &     SOFI   & 1999 Apr 10   \\
HM 15    &            11 07 44.5 & $-$77 39 43 &     SOFI   & 1999 Apr 10   \\
Cha H$\alpha$ 3 &            11 07 58.2 & $-$77 37 21 &     SOFI   & 1999 Apr 10   \\
SZ 23    &            11 07 58.4 & $-$77 42 42 &     SOFI   & 1999 Apr 10   \\
HM 16    &            11 07 59.1 & $-$77 38 46 &     SOFI   & 1999 Apr 10   \\
VW Cha   &            11 08 01.8 & $-$77 42 29 &     SOFI/OSIRIS  & 1999 Apr 10/1999 May 06  \\
Glass I  &            11 08 15.4 & $-$77 33 53 &     OSIRIS & 1999 May 06   \\
HM 19    &            11 08 16.9 & $-$77 44 37 &     OSIRIS & 1999 May 06   \\
Cha H$\alpha$ 4 &            11 08 20.6 & $-$77 39 19 &     SOFI   & 1999 Apr 10   \\
Cha H$\alpha$ 5 &            11 08 24.7 & $-$77 41 47 &     SOFI   & 1999 Apr 10   \\
Cha IR Nebula &       11 08 38.7 & $-$77 43 51 &     SOFI   & 1999 Apr 10   \\
Cha H$\alpha$ 6 &            11 08 39.6 & $-$77 34 17 &     SOFI   & 1999 Apr 10   \\
CHXR 78c &            11 08 54.2 & $-$77 32 12 &     OSIRIS & 1999 May 06   \\
HJM C 1-6 &           11 09 23.5 & $-$76 34 32 &     SOFI   & 1999 Apr 12   \\
HJM C 7-2 &           11 09 25.9 & $-$77 26 25 &     SOFI   & 1999 Apr 12   \\
HJM C 1-25 &          11 09 41.8 & $-$76 34 59 &     SOFI   & 1999 Apr 11   \\
HJM C 2-3  &          11 09 46.8 & $-$76 43 54 &     SOFI   & 1999 Apr 12   \\
Hn 10  &              11 09 47.5 & $-$76 34 45 &     SOFI   & 1999 Apr 11   \\
BYB 43 &              11 09 47.9 & $-$77 26 30 &     SOFI   & 1999 Apr 12   \\
HJM C 1-2 &           11 09 55.0 & $-$76 32 40 &     SOFI   & 1999 Apr 11   \\
HJM C 7-3 &           11 10 00.0 & $-$77 26 33 &     SOFI   & 1999 Apr 12   \\
Hn 11 &               11 10 04.9 & $-$76 33 28 &     SOFI   & 1999 Apr 11   \\
GK-1  &               11 10 04.9 & $-$76 35 47 &     SOFI   & 1999 Apr 11   \\
BYB 53 &              11 14 50.8 & $-$77 33 39 &     OSIRIS & 1999 May 07   \\
&   &  &   &   \\
\multicolumn{1}{l}{Lupus} \\   
&   &  &   &   \\
SZ 65 &               15 39 27.7 & $-$34 46 17 &     OSIRIS & 1999 May 09   \\
SZ 68 &               15 45 12.9 & $-$34 17 31 &     OSIRIS & 1999 May 06   \\
SZ 77 &               15 51 47   & $-$35 56 43 &     OSIRIS & 1999 May 08   \\
SZ 82 &               15 56 09.3 & $-$37 56 06 &     OSIRIS & 1999 May 08   \\
SZ 128 &              15 58 07.4 & $-$41 51 48 &     SOFI   & 1999 Apr 10   \\
HO Lup &              16 07 00.6 & $-$39 02 19 &     OSIRIS & 1999 May 09   \\
SZ 90  &              16 07 10.1 & $-$39 11 03 &     OSIRIS & 1999 May 07   \\
SZ 94  &              16 07 49.6 & $-$39 04 29 &     SOFI   & 1999 Apr 12   \\
SZ 96  &              16 08 12.6 & $-$39 08 33 &     OSIRIS & 1999 May 06   \\
HK Lup &              16 08 22.5 & $-$39 04 46 &     OSIRIS & 1999 May 06   \\        
SZ 103 &              16 08 30.3 & $-$39 06 11 &     OSIRIS & 1999 May 09   \\
Eggen 2 &             16 08 42.7 & $-$39 06 18 &     OSIRIS & 1999 May 06   \\
SZ 114  &             16 09 01.8 & $-$39 05 12 &     OSIRIS & 1999 May 08   \\
SZ 117  &             16 09 44.3 & $-$39 13 30 &     OSIRIS & 1999 May 08   \\
SZ 123  &             16 10 51.5 & $-$38 53 14 &     OSIRIS & 1999 May 08   \\
SZ 119  &             16 09 57.1 & $-$38 59 48 &     SOFI/OSIRIS & 1999 Apr 10/1999 May 07  \\
SZ 121  &             16 10 12.2 & $-$39 21 01 &     SOFI   & 1999 Apr 10   \\
SZ 122  &             16 10 16.4 & $-$39 08 00 &     OSIRIS & 1999 May 06   \\
SZ 124  &             16 11 53.3 & $-$39 02 16 &     SOFI/OSIRIS & 1999 Apr 10/1999 May 08  \\
&   &  &   &   \\
\multicolumn{1}{l}{Other clouds} \\   
&   &  &   &   \\
RX J0842.4-8345 &     08 42 22.4 & $-$83 45 24 &     SOFI & 1999 Apr 11    \\
RX J0844.8-7846 &     08 44 50.8 & $-$78 46 55 &     SOFI & 1999 Apr 11    \\
RX J0915.5-7608 &     09 15 30   & $-$76 08 00 &     SOFI & 1999 Apr 11    \\
RX J0951.9-7901 &     09 51 50.7 & $-$79 01 38 &     SOFI & 1999 Apr 11    \\     
Haro 1-1 &            16 21 34.4 & $-$26 12 25 &     OSIRIS & 1999 May 08  \\
ROX 13   &            16 26 29   & $-$24 19 42 &     OSIRIS & 1999 May 07  \\
Lkh$\alpha$ 104 &     18 02 54.3 & $-$24 20 56 &     OSIRIS & 1999 May 07  \\
RNO 93   &            17 16 13.8 & $-$20 57 46 &     OSIRIS & 1999 May 07  \\     
\enddata
\end{deluxetable}

\begin{deluxetable}{lrrrrl} 
\tabletypesize{\scriptsize}
\tablecolumns{6} 
\tablewidth{11cm}  
\tablecaption{Compiled photometry for the Chamaeleon I candidate young stellar objects} 
\tablehead{
\colhead{Name} & \colhead{$I$}   & \colhead{$J$}    & \colhead{$H$} & 
\colhead{$K$}    & \colhead{Reference}}
\startdata 

CCE98 1 &             11.09 & 9.08  & \nodata  & 7.50  & (1) \\
CCE98 2 &             10.65 & 8.93  & \nodata  & 7.47  & (1) \\   
CCE98 5 &             13.57 & 10.73 & \nodata  & 8.68  & (1) \\
CCE98 8 &           \nodata & 13.75 & \nodata  & 10.50 & (1) \\
CCE98 9 &           \nodata & 13.40 & \nodata  & 10.01 & (1) \\
CCE98 12 &            18.48 & 13.77 & 11.52    & 10.55 & (1), (2) \\
GK2001 3 &          \nodata & 14.26 & 13.23    & 12.48 & (2) \\
CCE98 14 &            18.13 & 13.65 & 11.46    & 10.57 & (1), (2) \\
CCE98 16 &            11.99 & 10.00 & \nodata  &  8.41 & (1) \\
CCE98 17 &            15.52 & 11.46 &  9.77    &  9.04 & (1), (2) \\
CCE98 18 &            17.98 & 13.32 & 10.96    &  9.92 & (1), (2) \\ 
CCE98 19 &            16.55 & 12.59 & 10.93    & 10.26 & (1), (2) \\ 
CCE98 20 &          \nodata & 16.47 & 14.19    & 13.22 & (1) \\ 
GK2001 15 &         \nodata & 12.65 & 11.72    & 10.22 & (2) \\
CCE98 21 &          \nodata & 14.66 & 12.46    & 11.27 & (2) \\
CCE98 23 &          \nodata & 17.55 & 13.60    & 10.97 & (2) \\
CCE98 24 &          \nodata & 15.20 & 12.42    & 10.83 & (2) \\
CCE98 25 &           12.32  &  9.46 & 8.09     &  7.63 & (1), (2) \\
CCE98 26 &           16.31  & 12.44 & 10.75    &  9.91 & (1), (2) \\
CCE98 27 &           18.03  & 13.60 & 11.73    & 10.91 & (1), (2) \\
GK2001 18 &         \nodata & 13.87 & 12.93    & 12.22 & (2) \\  
CCE98 32 &           14.39  & 11.51 &  9.67    &  8.30 & (1), (2) \\
CCE98 33 &          \nodata & 14.28 & 11.72    & 10.47 & (2) \\ 
CCE98 34 &          \nodata & 12.79 & 10.28    &  8.90 & (2) \\
GK2001 21 &         \nodata & 14.26 & 13.44    & 12.32 & (2) \\
GK2001 24 &         \nodata & 14.27 & 13.71    & 12.92 & (2) \\  
CCE98 35 &          \nodata & 15.52 & 13.86    & 12.56 & (2) \\ 
ISO-ChaI 177 &      \nodata & 13.02 & 12.20    & 11.56 & (2) \\    
CCE98 36 &          \nodata & 14.42 & \nodata  & 11.68 & (1) \\
CCE98 37 &            16.14 & 12.86 & 11.17    & 10.47 & (1), (2) \\   
CCE98 39 &            18.20 & 14.36 & 12.62    & 11.82 & (1), (2) \\
CCE98 40 &            17.77 & 13.15 & 11.19    & 10.17 & (1), (2) \\
PMK99 IR Cha INa1 &   \nodata & 18.24 & 15.28    & 12.75 & (3) \\
PMK99 IR Cha INa2 &   \nodata & 18.25 & 16.72    & 15.97 & (4) \\ 
CCE98 42 &            15.28 & 11.43 &  9.81    &  9.11 & (1), (2) \\
CCE98 43 &          \nodata & 15.56 & 13.26    & 12.29 & (2) \\
CCE98 44 &          \nodata & 15.36 & 13.09    & 11.86 & (2) \\
CCE98 46 &          \nodata & 15.66 & 13.38    & 12.42 & (2) \\  
CCE98 47 &            16.58 & 12.36 & 10.62    &  9.86 & (1), (2) \\
PMK99 IR Cha INa3 &   \nodata & 17.79 & 16.31    & 14.98 & (3) \\ 
PMK99 IR Cha INa4 &   \nodata & 17.47 & 15.92    & 14.36 & (3) \\
CCE98 48 &            13.90 & 10.75 &  9.34    &  8.55 & (1), (2) \\
CCE98 49 &            17.51 & 13.33 & 11.71  & 10.83 & (1), (2) \\
GK2001 42 &         \nodata & 14.58 & 13.94    & 13.25 & (2) \\
CCE98 50 &            12.78 &  9.88 & \nodata  &  7.99 & (1), (2) \\ 
CCE98 51 &            14.90 & 11.72 & 10.17    &  9.61 & (1), (2) \\ 
\enddata

\tablerefs{ 
(1): \cite{cam98}, (2): \cite{goke01}, (3): \cite{per99}, (4): \cite{oas99}.}
\end{deluxetable}

%% The following command ends your manuscript. LaTeX will ignore any text

\begin{deluxetable}{lrrrrlll}
\tabletypesize{\scriptsize}  
\tablecolumns{9}
\tablewidth{15cm}
\tablecaption{Previously known young stellar objects: Compiled photometry and optical spectral types}
\tablehead{
\colhead{Name} & \colhead{$I$} & \colhead{$J$}    &
\colhead{$H$} & \colhead{$K$} & \colhead{Type{$^{\rm a}$}} & \colhead{ST} &
\colhead{Referece} }  
\startdata                    

&   &  &   & & & &   \\
\multicolumn{1}{l}{Chamaeleon I} \\   
&   &  &   & & & &   \\
Hn 1    &             12.97 & 11.32 & 10.40 &  10.08 &      W            &  M2        & (1), (2), (13) \\
BYB 18 &             13.63 & 11.80 & 11.01 &  10.64 &       W            &  \nodata   & (1), (2), (3) \\ 
CHXR 73 &           \nodata & 12.60 & 11.24 &  10.70 &      W            &  K3-M2     & (2), (3) \\
Ced 110 IRS 4 &     \nodata & 16.53 & 13.88 &  12.57 &      \nodata      &  \nodata   & (3) \\
CHXR 74 &           \nodata & 11.55 & 10.55 &  10.25 &     \nodata       &  K7        & (2),(3),(4) \\
UY Cha  &             12.92 & 11.10 & 10.33 &   9.91 &      C            &  M1.5      & (1),(3),(5) \\
Ced 110 IRS 6 &     \nodata & \nodata & 14.61 & 10.81 &     \nodata      & \nodata    & (3) \\
Hn 6  &               13.34 & 11.11   & 10.07 &  9.72 &     W            &  K7        &  (1), (2), (3), (13) \\
Cha H$\alpha$ 1 &    16.4   & 13.42   & 12.76 & 12.31 &     C            & M7.5       & (3), (6)  \\
LkH$\alpha$ 332-17 &  9.31  &  7.99   &  7.21 &  6.77 &     C            & G2         & (1), (3), (5) \\
BYB 34 &             14.43  & 12.19   & 11.30 & 10.95 &     W            & M5         & (1), (2), (3), (4) \\ 
CHXR 26 &            15.16  & 11.41   &  9.88 &  9.21 &     \nodata      & K7         & (1), (2), (3) \\ 
Cha H$\alpha$ 2 &    15.3   & 12.21   & 11.22 & 10.65 &     C            & M6.5       & (3), (6) \\ 
HM 15    &           12.56  & 10.22   &  9.05 &  8.42 &     C            & M0.5       & (1), (3), (5) \\
Cha H$\alpha$ 3 &    15.0   & 12.35   & 11.69 & 11.27 &     W            & M7         & (3), (6) \\ 
SZ 23    &           14.58  & 11.85   & 10.62 &  9.81 &     C            & M2.5       & (1), (3), (4) \\
HM 16    &           12.76  &  9.71   &  8.19   & 7.16 &    C            & K7         & (1), (3), (4), (5) \\ 
VW Cha   &           10.56  &  8.63   &  7.72   & 7.17 &    C            & K2-K5      & (1), (3), (5), (14) \\
Glass I  &           10.54  &  8.58   &  7.40   & 6.22 &    W            & K4         & (1), (3), (5) \\
HM 19    &           13.04  & 11.29   & 10.41   & 10.12 &   W            & M3.5       & (1), (3) \\
Cha H$\alpha$ 4 &    14.4   & 12.13   & 11.43   & 11.08 &   W            & M6         & (3), (6) \\ 
Cha H$\alpha$ 5 &    14.7   & 12.13   & 11.21   & 10.78 &   W            & M6         & (3), (6) \\
Cha IR Nebula &      15.86  & 11.12   &  9.30   & 8.07  &   \nodata      & \nodata    & (1), (3) \\
Cha H$\alpha$ 6 &    15.1   & 12.37   & 11.53   & 11.04 &   C            & M7         & (3), (6) \\
CHXR 78c &           14.75  & 12.36   & 11.54   & 11.17 &   W            & M5.5       & (1), (2), (3), (4) \\
HJM C 1-6 &          17.99  & 12.36   & 10.13   &  8.52 &   \nodata      & \nodata    & (1), (3) \\
HJM C 7-2 &          \nodata & 11.64  & 10.05   &  9.34 &   \nodata      & \nodata    & (7) \\
HJM C 1-25 &         \nodata & 13.48  & 11.15   &  9.63 &   \nodata      & \nodata    & (3) \\ 
HJM C 2-3  &         14.50   & 11.73  & 10.72   & 10.14 &   \nodata      & \nodata    & (1), (3) \\
Hn 10  &             14.81   & 11.94  & 10.68   &  9.96 &   \nodata      & \nodata    & (1), (3), (13) \\
BYB 43 &             16.02   & 12.61  & 11.12   & 10.12 &   \nodata      & \nodata    & (1), (3) \\
HJM C 1-2 &          \nodata & 13.42  & 10.98   &  9.25 &   \nodata      & \nodata    & (3) \\ 
HJM C 7-3 &          \nodata & 12.11  & 10.51   &  9.92 &   \nodata      & \nodata    &  (7) \\ 
Hn 11 &               14.46  & 11.53  & 10.05   &  9.20 &   C            & \nodata    & (1), (3), (13) \\
GK-1  &               12.86  & 10.39  &  9.48   &  9.10 &   W            &  M0        & (3), (5), (14) \\ 
BYB 53 &              11.89  & 10.49  &  9.76   &  9.53 &   W            &  M1        & (1), (2), (3) \\ 
&   &  &   &  &  &  \\
\multicolumn{1}{l}{Lupus} \\   
&   &  &   &  &  &  \\
SZ 65 &              10.51  & 9.19  &  8.39   &  7.99 & C      & M0-M0.5    & (8), (14) \\ 
SZ 68 &               8.86  & 7.65  &  6.94   &  6.55 & W      & K2         & (8) \\ 
SZ 77 &              10.83  & 9.53  &  8.72   &  8.34 & C      & M0         & (8), (14) \\
SZ 82 &              10.20  & 8.87  &  8.10   &  7.75 & W      & M0         & (8), (14) \\ 
SZ 128 &             12.70  & 11.19 & 10.29   &  9.71 & W      & M1.5       & (8) \\ 
HO Lup &             11.42  & 10.04 & 9.14    &  8.60 & C      & M1         & (8)  \\
SZ 90  &             13.28  & 10.46 & 9.44    &  8.82 & C      & K7-M0      & (8), (14)  \\
SZ 94  &             12.90  & 11.50 & 10.89   & 10.63 & W      & M4         & (8) \\ 
SZ 96  &             11.84  & 10.21 & 9.35    & 8.90  & C      & M1.5       & (8), (14)  \\         
HK Lup &             10.92  &  9.55 & 8.64    & 7.99  & C      & K7-M0      & (8), (14)  \\ 
SZ 103 &             13.37  & 11.78 & 10.90   & 10.34 & C      & M4         & (8), (14) \\ 
Eggen 2 &            11.16  &  9.72 &  8.98   &  8.75 & W      & M1         & (8) \\ 
SZ 114  &            12.12  & 10.42 &  9.62   &  9.18 & C      & M4         & (8) \\ 
SZ 117  &            12.27  & 11.02 &  10.11  &  9.66 & C      & M2         & (8) \\
SZ 123  &            12.78  & 11.20 &  10.36  &  9.91 & C      & M1         & (8) \\ 
SZ 119  &            11.96  & 10.36 &  9.68   &  9.45 & W      & M4         & (8)  \\
SZ 121  &            11.84  & 10.27 &  9.49   &  9.13 & W      & M3         & (8)  \\
SZ 122  &            12.12  & 11.03 & 10.31   & 10.07 & W      & M2         & (8)  \\ 
SZ 124  &            11.46  & 10.38 &  9.79   &  9.48 & W      & K7-M0      & (8) \\
&   &  &   &  & & &  \\
\multicolumn{1}{l}{Other clouds} \\   
&   &  &   &  & & &  \\
RX J0842.4-8345 &    10.53  & 9.44    &  8.82   &  8.65   & W           & K4-K6     & (9), (10) \\ 
RX J0844.8-7846 &    10.70  & 9.61    &  8.93   &  8.72   & W           & K6-M0     & (9), (10) \\
RX J0915.5-7608 &    10.35  & 9.37    &  8.71   &  8.52   & W           & K6-K7     & (9), (10) \\ 
RX J0951.9-7901 &     9.29  & 8.66    &  8.20   &  8.09   & W           & G7-G8     & (9), (10) \\ 
Haro 1-1 &          \nodata & \nodata & \nodata & \nodata & C           & K5-K7     & (11) \\ 
ROX 13   &          \nodata & \nodata & \nodata & \nodata & \nodata     & \nodata   & \\ 
Lkh$\alpha$ 104 &   \nodata & \nodata & \nodata & \nodata & C           & K5        &  (12)\\  
RNO 93   &          \nodata & \nodata & \nodata & \nodata & C           & K5        &  (12) \\ 

\enddata

\tablenotetext{a}{C and W refer to the classical and weak emission T Tauri type, 
(CTTS and WTTS)} 

\tablerefs{
(1): \cite{cam98}, (2): \cite{law96},
(3): \cite{goke01}, (4): \cite{com99}, (5): \cite{gast92},
(6): \cite{neco99}, (7): \cite{kego01}, (8): \cite{hug94},
(9): \cite{alc95}, (10): \cite{cov97}, (11): \cite{shhe98},
(12): \cite{coku79}, (13): \cite{har93} (14): \cite{app83}.}    
\end{deluxetable} 

\begin{deluxetable}{lrrrrr} 
\tabletypesize{\scriptsize}
\tablecolumns{6} 
\tablewidth{10cm}  
\tablecaption{Equivalent widths (in \AA) for the candidate young stars} 
\tablehead{
\colhead{Name $^{\mathrm{a}}$} & \colhead{ Pa$\beta$}   & \colhead{Br$\gamma$}    & \colhead{Na I$^{\mathrm{b}}$} & 
\colhead{Ca I$^{\mathrm{c}}$}    & \colhead{CO$^{\mathrm{d}}$}}

\startdata 

CCE98 1 &                 \nodata &    2    &   6      & 4       &  46     \\
CCE98 2 &                 \nodata & \nodata &   2      & 3       &  35     \\
CCE98 5 &                 \nodata &    2    &   3      & \nodata &  28     \\
CCE98 8 &                 \nodata &    2    &   3      & 2       &  18     \\
CCE98 9 &                 \nodata & \nodata &   3      & \nodata &   8     \\ 
CCE98 12 &                \nodata &    1    &   6      & 4       &  34     \\
CCE98 14 &                \nodata & \nodata &   3      & 3       &  28     \\
CCE98 16 &                \nodata &    2    &   4      & 3       &  39     \\
CCE98 17 &                \nodata &    2    & \nodata  & 1       &  15     \\
CCE98 18 &                \nodata &    2    &   2      & 2       &  22     \\     
CCE98 19 &                \nodata & \nodata &   5      & 5       &  20     \\
CCE98 21 &                \nodata &    1    &   4      & 5       &  19     \\
CCE98 23 &                \nodata & \nodata &   3      & 3       &  11     \\
CCE98 24 &                $-$8    &  $-$4   &   3      & 4       &  12     \\
CCE98 25 &                \nodata &     3   &   6      & \nodata &  65     \\ 
CCE98 32 &                \nodata & \nodata &   5      & \nodata &  \nodata \\
CCE98 33 &                \nodata &     2   &   1      & 1       &  24     \\
CCE98 34 &                \nodata & \nodata &   4      & 2       &  23     \\
CCE98 36 &                \nodata &     1   & \nodata  & \nodata &  19     \\
CCE98 37 &                \nodata & \nodata &   5      & \nodata &  26     \\
CCE98 40 &                \nodata &     7   & \nodata  & \nodata &  \nodata \\ 
PMK99 IR Cha INa1 &       \nodata & \nodata & \nodata  & \nodata &  \nodata \\
CCE98 42 &                \nodata & \nodata &   7      &  3      &   23   \\
CCE98 43 &                \nodata & \nodata &   4      & \nodata &   21     \\
CCE98 44 &                \nodata &     8   & \nodata  & \nodata &  \nodata  \\ 
CCE98 47 &                \nodata &     2   & \nodata  & \nodata &   17     \\
CCE98 48 &                \nodata & \nodata &   3      &  3      &   15     \\
CCE98 50 &                \nodata & \nodata &   5      &  2      &   35     \\
CCE98 51 &                \nodata &     2   &   3      & \nodata &  \nodata \\ 
\enddata

\tablenotetext{a}{Sources GK2001 3, CCE98 20, GK2001 15, GM 1,
CCE98 26, CCE98 27, GK2001 18, GK2001 21, GK2001 24, CCE98 35, ISO-ChaI 177,
CCE98 39, PMK99 IR Cha INa2, CCE98 46,
PMK99 IR Cha INa3, PMK99 IR Cha INa4, CCE98 49, and GK2001 42
were not measured due to poor S/N on the corresponding spectrum.} 

\tablenotetext{b}{Na I doublet (2.206 and 2.209 $\mu$m)}
 
\tablenotetext{c}{Ca I triplet (2.261, 2,263, and 2.266 $\mu$m)}
 
\tablenotetext{d}{CO $\nu$ $=$ 2-0, 3-1} 
 
\tablecomments{
Positive values indicate absorptions and negative values correspond
to emissions. Missing equivalent widths indicate that the corresponding feature
was not detected either in absorption or emission, at the resolution used and
given the the S/N ratio for each spectrum.}

\tablecomments{For the SOFI data we estimate an uncertainty of $\sim$ 1-2 \AA~ in the
Pa$\beta$, Br$\gamma$,  Na I doublet (2.206 and 2.209 $\mu$m), and Ca I triplet
(2.261, 2,263, and 2.266 $\mu$m) equivalent widths. The combined
CO $\nu$ $=$ 2-0 and 3-1 bands have a slightly worse precision of
$\sim$ 3 \AA.} 

\tablecomments{For the OSIRIS data we estimate an error of 2-3 \AA~ in
our measurements of Pa$\beta$, Br$\gamma$,  Na I doublet, and Ca I triplet.
The combined CO $\nu$ $=$ 2-0, and 3-1 bands have a precision of 
$\sim$ 4-5~ \AA.}

\end{deluxetable} 

\begin{deluxetable}{lrrrrrll}
\tabletypesize{\scriptsize}  
\tablecolumns{6}
\tablewidth{10cm}
\tablecaption{Equivalent widths (in \AA) for the observed previously known pre-main sequence stars}
\tablehead{\colhead{Name} & \colhead{ Pa$\beta$}   & \colhead{Br$\gamma$}  & \colhead{Na
I$^{\mathrm{a}}$} & \colhead{Ca I$^{\mathrm{b}}$}    & \colhead{CO$^{\mathrm{c}}$} &
\colhead{Note}}

% 1.28$\mu$m &  2.17$\mu$m & 2.21 $\mu$m & 2.26 $\mu$m & 2.23--2.38 $\mu$m & \\   

\startdata                    
&   &  &   & &     \\
\multicolumn{1}{l}{Chamaeleon I} \\   
&   &  &   & &     \\
Hn 1    &           \nodata &     \nodata &   3        &    2     &    8    & \\
BYB 18  &           \nodata &     \nodata &   7        &    4     &    33   & \\  
CHXR 73 &           \nodata &        2    &   \nodata  &  \nodata &    16   & \\
Ced 110 IRS 4 &     \nodata &     \nodata &   \nodata  &  \nodata & \nodata & \\ 
CHXR 74 &           \nodata &     \nodata &   6        &    5     &    13   & \\
UY Cha  &            $-$5   &      $-$2   &   4        &    4     &    15   & \\
Ced 110 IRS 6 &     \nodata &      $-$7   &   \nodata  &  \nodata & \nodata & \\
Hn 6  &             \nodata &     \nodata &   5        &    4     &    10   & \\
Cha H$\alpha$ 1 &   \nodata &     \nodata &   \nodata  &  \nodata &    13   & \\
LkH$\alpha$ 332-17 & $-$5   &     \nodata &   \nodata  &  \nodata & \nodata & \\
BYB 34 &            \nodata &     \nodata &   5        &    6     &    12   & \\ 
CHXR 26 &           \nodata &     \nodata &   9        &  \nodata &  \nodata & \\
Cha H$\alpha$ 2 &   \nodata &     \nodata &   5        &    6     &    12   & \\ 
HM 15    &           $-$5   &     \nodata &   4        &    2     &     8   & \\
Cha H$\alpha$ 3 &   \nodata &     \nodata &   3        &    2     &    14   & \\
SZ 23    &           $-$6   &      $-$2   &   6        &    5     &     9   & \\
HM 16    &           $-$8   &     \nodata &   \nodata  &  \nodata & \nodata & \\      
VW Cha   &           $-$8   &      $-$3   &   4        &    2     &     7   & SOFI data \\
VW Cha   &           $-$6   &      $-$2   &   5        &  \nodata &     6   & OSIRIS data \\
Glass I  &          \nodata &     \nodata &   \nodata  &  \nodata & \nodata & \\ 
HM 19    &          \nodata &     \nodata &   5        &    2     &    21   & \\
Cha H$\alpha$ 4 &   \nodata &     \nodata &   3        &    2     &    13   & \\ 
Cha H$\alpha$ 5 &   \nodata &     \nodata &   4        &    2     &    13   & \\
Cha IR Nebula &     \nodata &     \nodata &   2        &    2     &    16   & \\
Cha H$\alpha$ 6 &   \nodata &     \nodata &   3        &    2     &    10   & \\
CHXR 78c &          \nodata &        3    &   3        &  \nodata & \nodata & \\          
HJM C 1-6 &          $-$10  &     $-$2    &   1        &    2     & \nodata & \\   
HJM C 7-2 &         \nodata &        2    &   1        &    1     &    16   & \\
HJM C 1-25 &        \nodata &     \nodata &   3        &    3     &    7    & \\
HJM C 2-3  &         14     &       12    &   \nodata  &  \nodata & \nodata & \\
Hn 10  &             $-$3   &        3    &   4        &    4     &    9    & \\
BYB 43 &             $-$17  &      $-$4   &   3        &    2     &    9    & \\  
HJM C 1-2 &         \nodata &     \nodata &   2        &    2     &    5    & \\         
HJM C 7-3 &         \nodata &         2   &   3        &    2     &   19    & \\
Hn 11 &              $-$9   &      $-$4   &   1        &    3     &   7     & \\
GK-1  &              $-$2   &         1   &   6        &    4     &   23    & \\
BYB 53 &            \nodata &         2   &   3        &    3     &    8    & \\
&   &  &   &  &  \\
\multicolumn{1}{l}{Lupus} \\   
&   &  &   &  &  \\
SZ 65 &         \nodata &     \nodata &       2       &  \nodata &     6    &  \\
SZ 68 &         \nodata &        2    &       5       &    3     &    16    &  \\     
SZ 77 &         \nodata &        2    &       2       &    3     &    11    &  \\    
SZ 82 &         \nodata &     \nodata &       3       &    2     &     4    &  \\     
SZ 128 &        $-$2    &     \nodata &       5       &    5     &    12    &  \\   
HO Lup &        $-$12   &     $-$2    &       \nodata &  \nodata &     4    & \\     
SZ 90  &        $-$8    &     \nodata &       \nodata &    2     &  \nodata &  \\    
SZ 94  &        \nodata &     $-$6    &       10      &    4     &    17    &   \\    
SZ 96  &         $-$2   &     \nodata &       3       &  \nodata &    1     &  \\  
HK Lup &        \nodata &     \nodata &       3       &  \nodata &    7     &  \\     
SZ 103 &        \nodata &        3    &       4       &    7     &    19    &  \\
Eggen 2 &       \nodata &        4    &       4       &    6     &  \nodata & \\
SZ 114  &       \nodata &     \nodata &       4       &  \nodata &    3     & \\   
SZ 117  &       \nodata &     \nodata &       3       &  \nodata &    3     &  \\   
SZ 123  &       $-$8    &     \nodata &       2       &     3    &    2     & \\   
SZ 119  &       \nodata &     \nodata &       6       &     4    &    15    & SOFI data   \\ 
SZ 119  &       \nodata &     \nodata &       4       &     7    &    13    & OSIRIS data   \\         
SZ 121  &       \nodata &     \nodata &       6       &     6    &    15    & \\
SZ 122  &       \nodata &     \nodata &       4       &     5    &    5     & \\ 
SZ 124  &       \nodata &     \nodata &       6       &     6    &    12    & SOFI data   \\ 
SZ 124  &       \nodata &     \nodata &       5       &     6    &    8     & OSIRIS data   \\  
&   &  &   &  &   \\
\multicolumn{1}{l}{Other clouds} \\   
&   &  &   &  & \\
RX J0842.4-8345 &  \nodata &  \nodata &       6       &     5    &   8  & \\
RX J0844.8-7846 &  \nodata &  \nodata &       5       &     5    &   11 & \\
RX J0915.5-7608 &  \nodata &  \nodata &       5       &     5    &   11 & \\
RX J0951.9-7901 &  \nodata &  \nodata &       3       &     3    &   5  & \\
Haro 1-1 &          $-$16 &    $-$11  &       2       &     2    & \nodata  & \\
ROX 13   &         \nodata &     2    &       5       &     6    &   11  & \\
Lkh$\alpha$ 104 &  \nodata &     2    &   \nodata     &  \nodata & \nodata  & \\      
RNO 93   &         \nodata &  \nodata &       4       &     3    &   11  & \\
\enddata

\tablenotetext{b}{Na I doublet (2.206 and 2.209 $\mu$m)}
 
\tablenotetext{c}{Ca I triplet (2.261, 2,263, and 2.266 $\mu$m)}
 
\tablenotetext{d}{CO $\nu$ $=$ 2-0, 3-1}
 
\tablecomments{
Positive values indicate absorptions and negative values correspond
to emissions. Missing equivalent widths indicate that the corresponding feature
was not detected either in absorption or emission, at the resolution used and
given the the S/N ratio for each spectrum.}
 
\tablecomments{For the SOFI data we estimate an uncertainty of $\sim$ 1-2 \AA~
in the
Pa$\beta$, Br$\gamma$,  Na I doublet (2.206 and 2.209 $\mu$m), and Ca I triplet
(2.261, 2,263, and 2.266 $\mu$m) equivalent widths. The combined
CO $\nu$ $=$ 2-0 and 3-1 bands have a slightly worse precision of
$\sim$ 3 \AA.}
 
\tablecomments{For the OSIRIS data we estimate an error of 2-3 \AA~ in
our measurements of Pa$\beta$, Br$\gamma$,  Na I doublet, and Ca I triplet.
The combined CO $\nu$ $=$ 2-0, and 3-1 bands have a precision of
$\sim$ 4-5~ \AA.}         

\end{deluxetable} 

\begin{deluxetable}{llllllll} 
\tabletypesize{\scriptsize}
\tablecolumns{9} 
\tablewidth{18cm}  
\tablecaption{Spectral types and derived parameters for the Chamaeleon I candidate young stellar objects} 
\tablehead{\colhead{Name} & \colhead{$Q${$^{\rm a}$}} & \colhead{$\rm I_{H_2O}${$^{\rm b}$}} & \colhead{$\rm A_{J}$} &
\colhead{$\rm r_K$} & \colhead{$\rm Log (L_{bol}/L_{\sun})$} & \colhead{$\rm M/M_{\sun}$} & \colhead{Age [yr] }}
\startdata 

CCE98 1 &            \nodata & M0      & 1.2     & \nodata  & $+$0.52      & 0.28   & $<$ 1 $\times$ 10$^5$ \\
CCE98 2 &            \nodata & M0      & 1.0     & \nodata  & $+$0.50      & 0.28   & $<$ 1 $\times$ 10$^5$ \\   
CCE98 5 &            \nodata & M1      & 2.0     & \nodata  & $+$0.15      & 0.275  & 1 $\times$ 10$^5$ \\
CCE98 8 &            \nodata & M1      & 3.9     & \nodata  & $-$0.26   & 0.38   & 8 $\times$ 10$^5$ \\
CCE98 12 &           \nodata & M0      & 4.1     & $-$0.26  & $-$0.19   & 0.40   & 8 $\times$ 10$^5$  \\
GK2001 3 &           \nodata & \nodata & \nodata &  \nodata & \nodata   & \nodata & \nodata  \\
CCE98 14 &           \nodata & M0.5    & 4.0     & $+$0.33     & $-$0.20   & 0.38    &  7 $\times$ 10$^5$ \\
CCE98 16 &           \nodata & M1      & 1.2     &  \nodata & $+$0.14      & 0.275   & 1 $\times$ 10$^5$ \\
CCE98 17 &                M0 & M0.5    & 2.7     & $-$0.15  & $+$0.15      & 0.295   & 1 $\times$ 10$^5$ \\
CCE98 18 &              M1.5 & M0.5    & 4.4     & $-$0.28  & $+$0.11      & 0.275   & 1 $\times$ 10$^5$ \\ 
CCE98 19 &              M3   & M2      & 2.9     & $-$0.27  & $-$0.31   & 0.17    & 1 $\times$ 10$^5$ \\ 
CCE98 21 &                M3 & M2.5    & 4.3     & $-$0.18  & $-$0.60   & 0.18    & 7.5 $\times$ 10$^5$ \\
CCE98 23 &              M0.5 & M1      & 8.6     & $+$0.1   & $+$0.10    & 0.275   & 1 $\times$ 10$^5$ \\
CCE98 24 &              M2   & M2      & 5.9     & $-$0.2   & $-$0.15 & 0.22    & 1 $\times$ 10$^5$ \\
CCE98 25 &           \nodata & M0.5    & 1.8     & $-$0.2   & $+$0.61    & 0.25    & $<$ 1 $\times$ 10$^5$ \\
CCE98 26 &           \nodata & \nodata & \nodata & \nodata  & \nodata   & \nodata & \nodata \\
CCE98 27 &           \nodata & \nodata & \nodata & \nodata  & \nodata   & \nodata &  \nodata \\
GK2001 18 &          \nodata & M6.5    & 0.8     & $+$0.1   & $-$1.74   & 0.04    & 5 $\times$ 10$^5$ \\  
CCE98 32 &           \nodata & M2      & 3.4     & $+$0.3   & $+$0.34      & 0.17    & $<$ 1 $\times$ 10$^5$ \\
CCE98 33 &           \nodata & M0.5    & 4.9     & $-$0.1   & $-$0.06   & 0.34    & 4 $\times$ 10$^5$ \\ 
CCE98 34 &           \nodata & M1      & 4.8     & $-$0.1   & $+$0.48      & 0.235   & $<$ 1 $\times$ 10$^5$ \\
GK2001 21 &          \nodata & \nodata & \nodata & \nodata  & \nodata   & \nodata & \nodata \\
GK2001 24 &          \nodata & \nodata & \nodata & \nodata  & \nodata   & \nodata & \nodata \\  
ISO-ChaI 177 &      \nodata & M3      & 0.7     & $+$0.2    & $-$1.40   & 0.225   & 8 $\times$ 10$^6$ \\    
CCE98 36 &          \nodata & M0.5    & 3.1     & \nodata  & $-$0.86   & 0.60    & 1.5 $\times$ 10$^7$ \\ 
CCE98 37 &          \nodata & M1      & 2.7     & $-$0.2   & $-$0.41   & 0.41    & 1.5 $\times$ 10$^6$ \\   
PMK99 IR Cha INa1{$^{\rm c}$} &  \nodata & M3.5(M6.5)  & 6.3(6.15)   &    0.7(0.52)   & $-$1.25($-$1.36)   & 0.19(0.06) &
4 $\times$ 10$^6$(4 $\times$ 10$^5$) \\
PMK99 IR Cha INa2 &   \nodata & \nodata & \nodata & \nodata  & \nodata   & \nodata & \nodata \\ 
CCE98 42 &            \nodata & M2      & 2.9     & $-$0.2   & $+$0.16      & 0.185   & $<$ 1 $\times$ 10$^5$ \\
CCE98 43 &                 M2 & M0.5    & 4.3     & $-$0.3   & $-$0.85   & 0.37    & 4 $\times$ 10$^6$ \\
CCE98 47 &                 M0 & M0      & 2.8     & $-$0.1   & $-$0.16   & 0.40    & 7 $\times$ 10$^5$ \\
PMK99 IR Cha INa3 &  \nodata & \nodata   & \nodata & \nodata  & \nodata   & \nodata & \nodata \\ 
PMK99 IR Cha INa4 &  \nodata & \nodata   & \nodata & \nodata  & \nodata   & \nodata & \nodata \\
CCE98 48 &              M0.5 & M1.5      & 1.9     & $+$0.1      & 0.125     & 0.275   & 1 $\times$ 10$^5$ \\
CCE98 49 &                M4 & M5        & 2.7     & $-$0.1 & $-$0.77    & 0.12    & 2.5 $\times$ 10$^5$ \\
GK2001 42 &         \nodata & \nodata   & \nodata & \nodata  & \nodata   & \nodata & \nodata \\ 
CCE98 50 &          \nodata & M2.5      & 1.9     &  \nodata & $+$0.35      & 0.155   & $<$ 1 $\times$ 10$^5$ \\
CCE98 51 &          \nodata & M0.5      & 2.3     &  $-$0.2  & $-$0.10   & 0.35    & 5 $\times$ 10$^5$ \\ 

\enddata

\tablenotetext{a}{$Q$-index-derived M{$_{\rm sub-type}$}.}

\tablenotetext{b}{${\rm I_{H_2O}}$-index-derived M{$_{\rm sub-type}$}.}

\tablenotetext{c}{Values in brackets are corrected due to the high veiling of this object, assuming
a M6 spectral type.} 

\end{deluxetable} 

\begin{deluxetable}{llllllll}
\tabletypesize{\scriptsize}  
\tablecolumns{8}
\tablewidth{13cm}
\tablecaption{Previously known young stellar objects: Near-infrared spectral types and derived parameters}
\tablehead{
\colhead{Name} & \colhead{$Q$}   & \colhead{$\rm I_{H_2O}$} & \colhead{$\rm
A_{J}$}
& \colhead{$\rm r_K$} & \colhead{$\rm Log (L_{bol}/L_{\sun})$} &
\colhead{$\rm M/M_{\sun}$} & \colhead{Age [yr] }}        
\startdata                    
BYB 18 &            \nodata &  M4   & 0.5     & 0.0     & $-$0.99  & 0.155    &      2 $\times$ 10$^6$ \\  
Cha IR Nebula &     M4.5    &  M5.5 & 3.2     & 0.0     & $+$0.32  & 0.11     &  $<$ 1 $\times$ 10$^5$ \\
HJM C 1-6 &         \nodata &  M1   & 4.1     & $+$0.3  & $+$0.36  & 0.205    &  $<$ 1 $\times$ 10$^5$ \\
HJM C 7-2 &          M0.5   &  M0.5 & 2.4     & $-$0.1  & $-$0.03  & 0.385    &      7 $\times$ 10$^5$ \\
HJM C 1-25 &         M2     &  M1.5 & 4.4     & $+$0.1  & $+$0.015 & 0.19     &  $<$ 1 $\times$ 10$^5$ \\
Hn 10      &         M3     &  M3   & 1.8     &    0.0  & $-$0.50  & 0.17     &      3 $\times$ 10$^5$ \\
BYB 43 &             M1.5   &  M3   & 2.5     & $+$0.1  & $-$0.50  & 0.17     &      3 $\times$ 10$^5$ \\
HJM C 1-2 &          \nodata&  M1.5 & 4.7     & $+$0.3  & $+$0.15  & 0.225    &  $<$ 1 $\times$ 10$^5$ \\
HJM C 7-3 &          M0     &  M0.5 & 2.4     & $-$0.2  & $-$0.21  & 0.345    &      4 $\times$ 10$^5$ \\
Hn 11 &             \nodata &  M1.5 & 2.1     & $+$0.1  & $-$0.10  & 0.265    &      2 $\times$ 10$^5$ \\ 
ROX 13   &          \nodata &  M1.5 & \nodata & \nodata & \nodata  & \nodata  & \nodata \\
\enddata
\end{deluxetable}

\begin{deluxetable}{lrllll}
\tabletypesize{\scriptsize}  
\tablecolumns{6}
\tablewidth{11cm}
\tablecaption{Previously known young stellar objects: Derived parameters}
\tablehead{
\colhead{Name} & \colhead{$\rm A_{J}$} & \colhead{$\rm r_K$} &
\colhead{$\rm Log (L_{bol}/L_{\sun})$} & \colhead{$\rm M/M_{\sun}$} & \colhead{Age [yr] }}                     
\startdata                    
&   &  & & &   \\
\multicolumn{1}{l}{Chamaeleon I} \\   
&   &  & & &    \\
Hn 1               &   1.0     & $-$0.1  & $-$0.55  &  0.25    & 1 $\times$ 10$^6$ \\
CHXR 73            &   1.8     & $-$0.1  & $-$0.59  &  0.80    & 1.5 $\times$ 10$^7$\\
CHXR 74            &   0.9     & $-$0.1  & $-$0.55  &  0.72    & 7 $\times$ 10$^6$ \\
UY Cha             &   1.4     & $-$0.1  & $-$0.22  &  0.275   & 5 $\times$ 10$^5$ \\
Hn 6               &   1.0     & $-$0.05  & $-$0.33  & 0.60     &  2.5 $\times$ 10$^6$ \\
Cha H$\alpha$ 1    &   0.1     & 0.0     &  $-$1.88 & 0.021    & $<$ 1 $\times$ 10$^5$ \\
LkH$\alpha$ 332-17 &   1.2     & 0.1     &  $+$1.22    & 2.4     & 2 $\times$ 10$^6$ \\
BYB 34             &   0.8     & $-$0.15 &  $-$1.08  & 0.125    & 1.5 $\times$ 10$^6$ \\
CHXR 26            &   2.3     & $-$0.1  &  $+$0.06    & 0.45     & 4.5 $\times$ 10$^5$ \\ 
Cha H$\alpha$ 2    &   1.0     & 0.0     &  $-$1.02  & 0.08     & 2.5 $\times$ 10$^5$ \\ 
HM 15              &   1.3     & $+$0.1  &  $+$0.10    & 0.30     & 2 $\times$ 10$^5$ \\
Cha H$\alpha$ 3    &   0.1     & 0.0     &  $-$1.44  & 0.045    & 2 $\times$ 10$^5$ \\ 
SZ 23              &   1.8     & 0.1     &  $-$0.47  & 0.20     & 3 $\times$ 10$^5$ \\
HM 16              &   2.3     & $+$0.3     &  $+$0.73    & 0.35     & $<$ 1 $\times$ 10$^5$ \\ 
VW Cha             &   0.8     & $+$0.2     &  $+$0.61   & 0.70     & 2.5 $\times$ 10$^5$ \\
Glass I            &   1.5     & $+$0.8     &  $+$0.93   & 0.80     & 1 $\times$ 10$^5$ \\
HM 19              &   0.8     & $-$0.2  & $-$0.67  & 0.16     & 7 $\times$ 10$^5$ \\
Cha H$\alpha$ 4    &   0.2     & $-$0.1  & $-$1.29  & 0.08     & 7.5 $\times$ 10$^5$ \\ 
Cha H$\alpha$ 5    &   0.8     & $-$0.1  & $-$1.06  & 0.09     & 4 $\times$ 10$^5$ \\
Cha H$\alpha$ 6    &   0.6     & $-$0.1  & $-$1.26  & 0.05     & 2 $\times$ 10$^5$  \\
CHXR 78c           &   0.6     & $-$0.1  & $-$1.24  & 0.10     & 2 $\times$ 10$^6$ \\
GK-1               &   0.6     & $+$0.1     &  $-$0.25 & 0.43     & 9 $\times$ 10$^5$ \\ 
BYB 53             &   0.2     & 0.0     &  $-$0.47 & 0.45     & 2 $\times$ 10$^6$ \\
&   &  &  & &  \\
\multicolumn{1}{l}{Lupus} \\   
&   &  &  &  &  \\
SZ 65 &               0.3  & $+$0.1  &   $+$0.12 &  0.32  & 2   $\times$ 10$^5$   \\ 
SZ 68 &               0.6  & $+$0.15 &   $+$0.97 &  1.5  & 2   $\times$ 10$^5$   \\ 
SZ 77 &               0.3  & $+$0.1  &   $-$0.01 &  0.335 & 4   $\times$ 10$^5$   \\
SZ 82 &               0.2  & $+$0.1  &   $+$0.21 &  0.31  & 1   $\times$ 10$^5$   \\ 
SZ 128 &              0.6  & $+$0.2  &   $-$0.58 &  0.35  & 2   $\times$ 10$^6$   \\ 
HO Lup &              0.6  & $+$0.2  &   $-$0.11 &  0.33  & 5   $\times$ 10$^5$   \\
SZ 90  &              0.9  & $+$0.2  &   $-$0.16 &  0.40  & 8   $\times$ 10$^5$   \\
SZ 94  &              0.1  & $+$0.0  &   $-$1.06 &  0.16  & 2   $\times$ 10$^6$   \\ 
SZ 96  &              0.5  & $+$0.1  &   $-$0.23 &  0.28  & 6.5 $\times$ 10$^5$ \\         
HK Lup &              0.6  & $+$0.3  &   $+$0.09 &  0.33  & 3   $\times$ 10$^5$   \\ 
SZ 103 &              0.8  & $+$0.15 &   $-$0.89 &  0.15  & 1.5 $\times$ 10$^6$ \\ 
Eggen 2 &             0.2  & $+$0.0  &   $-$0.15 &  0.33  & 7   $\times$ 10$^5$   \\ 
SZ 114  &             0.6  & $+$0.0  &   $-$0.43 &  0.135 & 1   $\times$ 10$^5$   \\ 
SZ 117  &             1.0  & $+$0.0  &   $-$0.44 &  0.24  & 8   $\times$ 10$^5$   \\
SZ 123  &             0.45 & $+$0.1  &   $-$0.64 &  0.50  & 4   $\times$ 10$^6$   \\ 
SZ 119  &             0.3  & $-$0.1  &   $-$0.53 &  0.135 & 2   $\times$ 10$^5$   \\
SZ 121  &             0.6  & 0.0     &   $-$0.34 &  0.17  & 1.5 $\times$ 10$^5$ \\
SZ 122  &             0.5  & $-$0.1  &   $-$0.65 &  0.25  & 1.5 $\times$ 10$^6$ \\ 
SZ 124  &             0.0  & $+$0.1  &   $-$0.48 &  0.51  & 2.5 $\times$ 10$^6$ \\
&   &  &  & &  \\

\multicolumn{1}{l}{Other clouds} \\   

&   &  &  & &   \\
RX J0842.4-8345 &     0.1  &    0.0 &  $-$0.02   & 0.44  & 1.5 $\times$ 10$^6$ \\ 
RX J0844.8-7846 &     0.1  & $+$0.1 &  $+$0.11   & 0.48  & 4   $\times$ 10$^5$ \\
RX J0915.5-7608 &     0.0  & $+$0.1 &  $-$0.04   & 0.68  & 7.5 $\times$ 10$^5$\\ 
RX J0951.9-7901 &     0.1  &    0.0 &  $+$0.43   & 1.5  & 7   $\times$ 10$^6$ \\ 

\enddata

\end{deluxetable}

\end{document}